\documentclass[12pt]{article}
\usepackage{amssymb,amscd,array}
\catcode `\@=11
\@addtoreset{equation}{section}

\newtheorem{thm}{Theorem}[section]

\newtheorem{prop}[thm]{Proposition}

\def\qed{\blacksquare}
\newcommand{\be}{\begin{equation}}
\newcommand{\ee}{\end{equation}}
\newcommand{\bea}{\begin{eqnarray}}
\newcommand{\eea}{\end{eqnarray}}
\newcommand{\R}{\mathbb{R}}
\newcommand{\N}{\mathbb{N}}

\begin{document}
\begin{titlepage}
\begin{center}
{\bf \Large{Massive Gravity as a Quantum Gauge Theory\\}}
\end{center}
\vskip 1.0truecm
\centerline{D. R. Grigore, 
\footnote{e-mail: grigore@theor1.theory.nipne.ro, grigore@theory.nipne.ro}}
\vskip5mm
\centerline{Dept. of Theor. Phys., Inst. Atomic Phys.}
\centerline{Bucharest-M\u agurele, P. O. Box MG 6, ROM\^ANIA}
\vskip 1cm
\centerline{G. Scharf, 
\footnote{e-mail: scharf@physik.unizh.ch}}
\vskip5mm
\centerline{Inst. of Theor. Phys., Univ Z\"urich.}
\centerline{Wintherhurersts 190, Z\"urich, SWITZERLAND}

\vskip 2cm
\bigskip \nopagebreak
\begin{abstract}
\noindent
We present a new point of view on the quantization of the massive 
gravitational field, namely we use exclusively the quantum framework of 
the second quantization. The Hilbert space of the many-gravitons system
is a Fock space
${\cal F}^{+}({\sf H}_{\rm graviton})$
where the one-particle Hilbert space ${\sf H}_{graviton}$
carries the direct sum of two unitary irreducible representations of the 
Poincar\'e group corresponding to
 two particles of mass
$m > 0$
and spins $2$ and $0$, respectively.
This Hilbert space is canonically isomorphic to a space of the type
$Ker(Q)/Im(Q)$
where $Q$ is a gauge charge defined in an extension of the Hilbert space 
${\cal H}_{\rm graviton}$
generated by the gravitational field 
$h_{\mu\nu}$
and some ghosts fields 
$u_{\mu}, \tilde{u}_{\mu}$
(which are vector Fermi fields) and
$v_{\mu}$
(which are vector field Bose fields.) 

Then we study the self interaction of massive gravity in the causal
framework. We obtain a solution which goes smoothly to the zero-mass
solution
   of linear quantum gravity up to a term depending on the
bosonic
  ghost field. This solution depends on two real constants as
it should
  be; these constants are related to the gravitational
constant
 and
   the
  cosmological constant.
   In the second order of
the
 perturbation theory
  we do not need a Higgs field, in sharp
contrast to
 Yang-Mills theory.
\end{abstract}
\end{titlepage}

\section{Introduction}
The quantization of gravity is an old standing problem of quantum field
theory. The solution of this problem in full generality is a highly
non-trivial problem which seems to be extremely complicated. (See however
the papers of Ashtekar and collaborators \cite{As}, \cite{Thie}). In 
\cite{Gr3} and \cite{Gri1}-\cite{Gri4}, \cite{Sc2} this problem was addressed 
for the linear gravitational field 
of zero mass.
Among the pioneering works in this approach we mention \cite{Ti}, \cite{Fa}, 
\cite{OP}, \cite{Gu2}, \cite{KO1}, \cite{KO2}. Using the result of this
analysis many computations have been done in the literature (see \cite{CLM},
\cite{GS1}, \cite{Do}, \cite{Za}, \cite{Ve}). 
 
One possible way to perform the quantization of the asymptotic 
gravitational field is to
linearize the classical theory of gravitation using the so called Goldberg
variables \cite{Go}, \cite{Gu1} and then to apply straightforward
quantization of the resulting free field theory. Because of the gauge invariance of
the theory (which in this case is the invariance under general coordinates
transformations) one obtains a constrained system and one tries to use a
Bleuler-Gupta type formalism, that is to start with a Hilbert space endowed
with a sesquilinear non-degenerate form and select the physical states as a
subspace of the type
$Q_{A} \Phi = 0, \quad A = 1,\dots,N$.

A related idea is to extend the Fock space to an auxiliary Hilbert space 
${\cal H}^{gh}$ 
including some fictitious fields, called ghosts, and construct a gauge charge
(i.e. an operator $Q$ verifying $Q^{2} = 0$) such that the physical Hilbert
space is
${\cal H}_{phys} \equiv {\it Ker}(Q)/{\it Im}(Q)$
(see for instance \cite{KO2} and references quoted there).  As a result of this
procedure, it is asserted that the {\it graviton}, i.e. the elementary quantum
particle must be a massless spin $2$ particle. The construction of the
gauge charge relies heavily on classical field theory arguments, because one
tries to obtain for the quantum gauge transformations expressions of the same
type as the general coordinates transformations appearing in general
relativity. This invariance is then promoted to a quantum gauge invariance
which should be implemented by the commutator with the  gauge charge $Q$.

It is an interesting problem to consider the case of massive gravity. This
case was analyzed many times ago \cite{DV}, \cite{Va}. In \cite{DV} it
is argued that even the quantization of the massive spin $2$ field is
problematic in the sense that no smooth limit
$m \mapsto 0$
exists. Some recent interest on this problem exists \cite{AGS}, \cite{LPR},
\cite{RSS} ,\cite{GS2} and \cite{DKP}.

We will show here that one can perform the quantization in such
a way that this limit is smooth. One finds out that the massive graviton has
a scalar partner of the same mass $m$. The construction is done in the
spirit of \cite{Gr3}.

We also mention that a rigorous construction of the Hilbert space of the
many-gravitons system is indispensable for the construction of the
corresponding  $S$-matrix in perturbation theory in the sense of Bogoliubov.
This construction emphasizes the basic r\^ole of causality in quantum field
theory. We obtain a solution for the interaction
Lagrangian (the first-order chronological product) which goes smoothly for
$m \searrow 0$
into the solution appearing in \cite{Sc2}.

The solution we obtain, up to second order of the perturbation theory, coincides
with the result of the perturbative development of the Einstein-Hilbert
Lagrangian with cosmological constant, if we make the identification
$\Lambda = 2 m^{2}$ and use Goldberg variables (see the Conclusions). 
We remark that in the second order of the
perturbation theory we do not need a Higgs field as in the case of Yang-Mills
theory. For this reason it seems to be impossible to find our massive
spin 2 gauge theory by means of the conventional Higgs mechanism.

\newpage
\section{The Quantization of the Asymptotic \\ Massive Gravitational Field 
\label{grav}}

One defines the graviton as  a certain unitary irreducible
representation of the Poincar\'e group corresponding to zero mass and
helicity $2$. In the case of massive gravity one should use the representation
of positive mass $m$ and spin $2$. These representation can be explicitely
described using the formalism
 of Hilbert space bundles, as presented for
instance in \cite{Va}, ch. VI.7 thm 6.20.
 Let us denote by
${\rm H}^{(m)}_{\rm gr}$
the one-particle Hilbert space of the graviton of mass $m$. The ensemble
of many gravitons is usually described by the associated Fock space
$
{\cal F}_{\rm graviton} = {\cal F}^{(+)}({\rm H}^{(m)}_{\rm gr})
$
where the upper $+$ sign indicates that the gravitons are assumed to be
Bosons according the the well-known spin-statistics 
theorem. The Hilbert
 space
$
{\cal F}_{\rm graviton}
$
is not very suitable for the construction of the perturbative series of the
scattering matrix $S$ in the sense of Bogoliubov. The way out is to construct
a larger Hilbert space
$
{\cal H}
$
where unphysical degrees of freedom are present. In this Hilbert space a 
(gravitational) gauge charge $Q$ acts which should be chosen such that it
squares to zero
$Q^{2} = 0 $;
in this case it makes sense to consider the factor space
$
{\cal H}_{phys} \equiv {\it Ker}(Q)/{\it Im}(Q)
$
which should be canonically isomorphic to
$
{\cal F}_{\rm graviton}.
$

Let us describe this construction. We use in this paper the following 
notations. The upper hyperboloid of mass
$m \geq 0$
is by definition
$X^{+}_{m} \equiv \{p \in \R^{4} \vert \quad \Vert p\Vert^{2} = m^{2}\}$;
it is a Borel set with the Lorentz invariant measure
$d\alpha_{m}^{+}(p) \equiv {d{\bf p} \over 2\omega({\bf p})}$.
Here the conventions are the following:
$\Vert \cdot \Vert$
is the Minkowski norm defined by
$
\Vert p \Vert^{2} \equiv p\cdot p
$
and
$p\cdot q$
is the Minkowski bilinear form:
$
p\cdot q \equiv p_{0}q_{0} - {\bf p} \cdot {\bf q};
$
by
$\eta_{\mu\nu}$
we denote the corresponding flat Minkowski matrix with diagonal elements
$1,-1,-1,-1$.
If
${\bf p} \in \R^{3}$
we define
$\tau({\bf p}) \in X_{m}^{+}$
according to
$
\tau({\bf p}) \equiv (\omega({\bf p}),{\bf p}), \quad
\omega({\bf p}) \equiv \sqrt{{\bf p}^{2} + m^{2}}.
$

First we consider the zero mass case
$m = 0$
\cite{Gr3}, \cite{Sc2}. 
\begin{itemize}
\item
One generates the Hilbert space
${\cal H}$
by applying on the vacuum the fields
$
H_{\mu\nu}, u_{\rho}, \tilde{u}_{\rho}, \Phi
$
(the rigorous construction is based on the Borchers algebra); these fields are
of null mass:
\be
\partial^{2} H_{\mu\nu}(x) = 0 \quad
\partial^{2} u_{\rho}(x) = 0 \quad
\partial^{2} \tilde{u}_{\rho}(x) = 0 \quad
\partial^{2} \Phi(x) = 0
\label{eq=0}
\ee
\item
$H_{\mu\nu}$
is symmetric and traceless:
\be
H_{\mu\nu} = H_{\nu\mu} \quad
{H^{\mu}}_{\mu} = 0
\ee
\item
The field
$\Phi$
is scalar and
$
H_{\mu\nu}, u_{\rho}, \tilde{u}_{\rho}
$
have obvious tensor and vector properties
\item
The fields
$
H_{\mu\nu}, \Phi
$
are Bosons and
$
u_{\rho}, \tilde{u}_{\rho}
$
are Fermions
\item
The causal commutation relations of these fields are
\bea
\left[H_{\rho\sigma}(x),H_{\lambda\omega}(y)\right] = 
- {i\over 2} \left( \eta_{\rho\lambda} \eta_{\sigma\omega} + 
\eta_{\rho\omega} \eta_{\sigma\lambda} - 
{1\over 2} \eta_{\rho\sigma} \eta_{\lambda\omega}\right)
D_{0}(x-y) \times {\bf 1}
\nonumber \\
\{u_{\mu}(x),\tilde{u}_{\nu}(y)\} = i~\eta_{\mu\nu}~D_{0}(x-y) {\bf 1}
\nonumber \\
~[\Phi(x),\Phi(y)] = i~D_{0}(x-y) {\bf 1}
\label{causal=0}
\eea 
and the other (anti)commutators are zero; in particular all Bose fields commute
with all Fermi fields. Here
\be
D_{m}(x) = D_{m}^{(+}(x) + D_{m}^{(-)}(x)
\ee
is the Pauli-Jordan distribution of mass
$m \geq 0$
and
$D^{(\pm)}_{m}(x)$
are given by:
\be
D_{m}^{(\pm)}(x) \equiv \pm {i\over (2\pi)^{3/2}} \int_{X^{+}_{m}}
d\alpha^{+}_{m}(p) e^{\mp i p\cdot x}.
\label{D0}
\ee
\item
In this Hilbert space there exists a sesqui-linear form (not positively 
defined)
$<\cdot,\cdot>$
such that we have
\bea
H_{\mu\nu}^{\dagger} = H_{\mu\nu} \quad
u_{\mu}^{\dagger} = u_{\mu} \quad
\tilde{u}_{\mu}^{\dagger} = - \tilde{u}_{\mu} 
\quad
\Phi^{\dagger} = \Phi
\label{conjugate=0}
\eea
where by $\dagger$ we mean the adjoint with respect to
$<\cdot,\cdot>$
\item
The operator $Q$ is well defined through the relations
\be
Q \Omega = 0
\label{Q-vacuum}
\ee
\bea
[Q, H_{\mu\nu}] = - {i\over 2}  
\left(\partial_{\mu}u_{\nu} + \partial_{\nu}u_{\mu} 
- {1\over 2} \eta_{\mu\nu} \partial_{\rho}u_{\sigma} \right)
\quad
[Q, \Phi] = {i\over 2} \partial^{\rho} u_{\rho}
\nonumber \\
\{Q,u_{\mu}\} = 0,\quad
\{Q,\tilde{u}_{\mu}\} =   
i~\left(\partial^{\nu} H_{\mu\nu} + {1\over 2} \partial_{\mu}\Phi\right).
\label{Q=0}
\eea
\end{itemize}
In these conditions one can prove that:
\begin{itemize}
\item
The operator $Q$ is well defined; for this one has to check the validity
of the Jacoby identity:
\bea
~[ b(x), \{ f(y), Q \} ] + \{ f(y), [ {Q}, b(x) ] \} = 0
\label{jacobi}
\eea
where
$b$ and $f$ are generic Bose (resp. Fermi) fields.
\item
The following relations are verified:
\be
Q^{2} = 0
\label{square}
\ee
\be
{\cal U}_{g} Q = Q {\cal U}_{g}, \quad \forall g \in {\cal P}.
\label{UQ}
\ee

From (\ref{square}) we have
\be
Im(Q) \subset Ker(Q)
\label{im-ker}
\ee
so it makes sense to consider the factor space
$
Ker(Q)/Im(Q).
$
One can prove that the sesqui-linear form
$<\cdot,\cdot>$
induces a strictly positively defined scalar product on 
$
\overline{Ker(Q)/Im(Q)}
$
and we have a canonical isomorphism 
$
\overline{Ker(Q)/Im(Q)} \sim {\cal F}_{gravition}.
$
\end{itemize}
The preceding construction presented in detail in \cite{Gr3} justifies the
consideration of the auxiliary Hilbert space
${\cal H}$
as a lay-ground for the perturbation theory. The fields
$
u_{\rho}, \tilde{u}_{\rho}, \Phi
$
are called {\it ghost} fields and the operator $Q$ is the 
{\it gauge charge}. A simplification of the preceding formalism is the
consideration of the new field
\be
h_{\mu\nu} \equiv H_{\mu\nu} + {1\over 2} \eta_{\mu\nu} \Phi
\label{newH}
\ee
which is self-adjoint in the sense
\be
h_{\mu\nu}(x)^{\dagger} = h_{\mu\nu}(x)
\ee
but is not traceless anymore and the causal commutation relations are:
\be
\left[h_{\rho\sigma}(x),h_{\lambda\omega}(y)\right] = 
- {i\over 2} \left( \eta_{\rho\lambda} \eta_{\sigma\omega} + 
\eta_{\rho\omega} \eta_{\sigma\lambda} 
- \eta_{\rho\sigma} \eta_{\lambda\omega}\right)
~D_{0}(x-y) \times {\bf 1}.
\label{commutation-grav-H}
\ee

We can easily prove that the preceding 
definition of the gauge charge is 
equivalently described by (\ref{Q-vacuum}) and:
\bea
~[Q, h_{\mu\nu}] = - {i\over 2}  
\left(\partial_{\mu}u_{\nu} + \partial_{\nu}u_{\mu} 
- \eta_{\mu\nu} \partial_{\rho}u^{\rho} \right)
\nonumber \\
\{Q,u_{\mu}\} = 0,\quad
\{Q,\tilde{u}_{\mu}\} = i~\partial^{\nu} h_{\mu\nu}
\label{Q-h=0}
\eea
so one can consider that the Hilbert space
${\cal H}$
is generated by the fields
$
h_{\mu\nu}, u_{\rho}, \tilde{u}_{\rho}
$
with the properties described above.

We now turn to the massive gravitational field. One notices from the very
beginning that in the case
$m > 0$
the gauge charge defined by (\ref{Q-h=0}) does not square to zero anymore.
One can try to correct this feature as in the case of the massive vector
field (see for instance \cite{Sc2}) by introducing a new ghost field
$v_{\mu}$
which is a vector field. The one modifies the preceding scheme as follows:
\begin{itemize}
\item
One generates the Hilbert space
${\cal H}$
by applying on the vacuum the fields
$
h_{\mu\nu}, u_{\rho}, \tilde{u}_{\rho}, v_{\mu}
;
$
all these fields are of mass $m$:
\be
(\partial^{2} + m^{2}) h_{\mu\nu}(x) = 0 \quad
(\partial^{2} + m^{2})u_{\rho}(x) = 0 \quad
(\partial^{2} + m^{2})\tilde{u}_{\rho}(x) = 0 \quad
(\partial^{2} + m^{2}) v_{\mu}(x) = 0
\label{eq=m}
\ee
\item
$h_{\mu\nu}$
is symmetric:
\be
h_{\mu\nu} = h_{\nu\mu}
\ee
\item
The fields
$
h_{\mu\nu}, u_{\rho}, \tilde{u}_{\rho}, v_{\mu}
$
have obvious tensor and vector properties
\item
The fields
$
h_{\mu\nu}, v_{\mu}
$
are Bosons and
$
u_{\rho}, \tilde{u}_{\rho}
$
are Fermions
\item
The causal commutation relations of these fields are
\bea
\left[h_{\rho\sigma}(x),h_{\lambda\omega}(y)\right] = 
- {i\over 2} \left( \eta_{\rho\lambda} \eta_{\sigma\omega} + 
\eta_{\rho\omega} \eta_{\sigma\lambda} - 
\eta_{\rho\sigma} \eta_{\lambda\omega}\right)
D_{m}(x-y) \times {\bf 1}
\nonumber \\
\{u_{\mu}(x),\tilde{u}_{\nu}(y)\} = i~\eta_{\mu\nu}~D_{m}(x-y) {\bf 1}
\nonumber \\
~[v_{\mu}(x),v_{\rho}(y)] = {i\over 2}~D_{m}(x-y) {\bf 1}
\label{causal=m}
\eea 
and the other (anti)commutators are zero
\item
In this Hilbert space there exists a sesqui-linear form (not positively 
defined)
$<\cdot,\cdot>$
such that we have
\bea
h_{\mu\nu}^{\dagger} = h_{\mu\nu} \quad
u_{\mu}^{\dagger} = u_{\mu} \quad
\tilde{u}_{\mu}^{\dagger} = - \tilde{u}_{\mu} 
\quad
v_{\mu}^{\dagger} = v_{\mu}
\label{conjugate=m}
\eea
where by $\dagger$ we mean the adjoint with respect to
$<\cdot,\cdot>$
\item
The operator $Q$ is well defined through the relations(\ref{Q-vacuum}
)
and
\bea
~[Q, h_{\mu\nu}] = - {i\over 2}  
\left(\partial_{\mu}u_{\nu} + \partial_{\nu}u_{\mu} 
- \eta_{\mu\nu} \partial_{\rho}u_{\sigma} \right)
\quad
[Q, v_{\mu}] = - {i m\over 2} u_{\mu}
\nonumber \\
\{Q,u_{\mu}\} = 0,\quad
\{Q,\tilde{u}_{\mu}\} =   
i~\left(\partial^{\nu} h_{\mu\nu} + m v_{\mu}\right).
\label{Q=m}
\eea
\end{itemize}
In these conditions one can prove that the operator $Q$ is well defined
because of the validity of the Jacobi identity (\ref{jacobi}) and we also have
(\ref{square}) and (\ref{UQ}), so again it makes sense to consider the 
factor space
$
\overline{Ker(Q)/Im(Q)}.
$
One can prove in this case also that the sesqui-linear form
$<\cdot,\cdot>$
induces a strictly positively defined scalar product on this factor space.
However, in this case a modification of the zero mass scheme appears. The
one-particle Hilbert corresponding to 
$
\overline{Ker(Q)/Im(Q)} 
$
is
$
{\rm H}^{[m,2]} \oplus {\rm H}^{[m,0]}
$
i.e. it describes two particles of mass $m$, one of spins $2$ and one of
spin 0, respectively. In other words we have
$
\overline{Ker(Q)/Im(Q)} = {\cal F}_{gravition} \oplus {\cal F}_{scalar}
.
$
It seems impossible to construct a gauge structure such that the scalar
partner of the graviton is eliminated, so in this paper we will accept
that such a particle does exists.
It remains to be investigated whether the scalar partner of the
graviton with a tiny mass leads to phenomenological problems.

Sometimes it is convenient to generalize the expression of the new field 
(\ref{newH}) in the sense:
\be
h_{\mu\nu}^{(\alpha)} \equiv H_{\mu\nu} + {1\over 2}\alpha~\eta_{\mu\nu} \Phi
\label{H-alpha}
\ee
with
$\alpha \in \R^{*}$.
The causal commutation relations are for this field:
\be
\left[h_{\rho\sigma}^{(\alpha)}(x),h_{\lambda\omega}^{(\alpha)}(y)\right] = 
- {i\over 2} \left( \eta_{\rho\lambda} \eta_{\sigma\omega} + 
\eta_{\rho\omega} \eta_{\sigma\lambda} 
- {1 + \alpha^{2} \over 2}~\eta_{\rho\sigma} \eta_{\lambda\omega}\right)
D_{m}(x-y) \times {\bf 1}.
\label{commutation-grav-H-alpha}
\ee

We can prove that the definition of the gauge charge is equivalently 
described by (\ref{Q-vacuum}) and:
\be
[Q, h_{\mu\nu}^{(\alpha)}] = - {i\over 2}  
\left(\partial_{\mu}u_{\nu} + \partial_{\nu}u_{\mu} 
- {1 + \alpha \over 2}~\eta_{\mu\nu} \partial_{\rho}u^{\rho} \right),
\ee
\be
\{Q,u_{\mu}\} = 0,\quad
\{Q,\tilde{u}_{\mu}\} =   
i~\left(\partial^{\nu}h_{\mu\nu}^{(\alpha)} 
+ {1 - \alpha\over 4\alpha}~\partial_{\mu}h^{(\alpha)}
+ m v_{\mu}\right)
\ee
and
\be
[Q, v_{\mu}] = - {i m\over 2} u_{\mu}; 
\ee
here
\be
h^{(\alpha)} \equiv \eta^{\mu\nu}~h^{(\alpha)}_{\mu\nu}.
\ee

The choice (\ref{newH}) correspond to 
$\alpha = 1$.
Let us consider the choice
$\alpha = - 1$.
Then the preceding relations for
\be
\hat{h}_{\mu\nu} = h^{(-1)}_{\mu\nu}
\label{hhat}
\ee
become:
\be
\left[\hat{h}_{\rho\sigma}(x),\hat{h}_{\lambda\omega}(y)\right] = 
- {i\over 2} \left( \eta_{\rho\lambda} \eta_{\sigma\omega} + 
\eta_{\rho\omega} \eta_{\sigma\lambda} 
- \eta_{\rho\sigma} \eta_{\lambda\omega}\right)
 D_{m}(x-y) \times {\bf 1}.
\ee
\be
[Q, \hat{h}_{\mu\nu}] = - {i\over 2}  
\left(\partial_{\mu}u_{\nu} + \partial_{\nu}u_{\mu} \right),
\ee
\be
\{Q,u_{\mu}\} = 0,\quad
\{Q,\tilde{u}_{\mu}\} =   
i~\left(\partial^{\nu}\hat{h}_{\mu\nu} - {1 \over 2}~\partial_{\mu}\hat{h}
+ m v_{\mu}\right)
\ee
and
\be
[Q, v_{\mu}] = - {i m\over 2} u_{\mu}. 
\ee

This choice seems to appear naturally in the classical framework of gravity
with an non-zero cosmological constant, if one expands the metric
$g_{\mu\nu}$ around Minkowski background in the form 
\bea
g_{\mu\nu} = \eta_{\mu\nu} + \kappa \hat{h}_{\mu\nu}
\nonumber
\eea
(see the Conclusions).
 However, from the quantum point of
view the value of $\alpha$ is irrelevant: all choices are good for the 
description of the physical Hilbert space.

We remark also that the massless limit 
problem mentioned in \cite{DV} has a
very simple explanation according to the preceding observation: in \cite{DV}
one uses different values of the parameter $\alpha$ for the case
$m = 0$ 
and 
$m > 0$
respectively. The correct procedure is to use the same value of $\alpha$
in both cases.

The construction of observables can be done in the usual way.
We denote by ${\cal W}$ the linear space of all Wick monomials on the
Fock space
${\cal H}^{gh}$
i.e. containing the fields
$h_{\mu\nu}(x)$,
$u_{\mu}(x)$,
$\tilde{u}_{\mu}(x)$
and
$v_{\mu}(x)$.
If $M$ is such a Wick monomial, we define by
$gh_{\pm}(M)$
the degree in 
$\tilde{u_{\mu}}$ 
(resp. in 
$u_{\mu}$). 
The total degree of $M$ is 
\be
deg(M) \equiv gh_{+}(M) + gh_{-}(M).
\ee

The {\it ghost number} is, by definition, the expression:
\be
gh(M) \equiv gh_{+}(M) - gh_{-}(M).
\ee

If
$M \in {\cal W}$
let us define the operator:
\be
d_{Q} M \equiv :QM: - (-1)^{gh(M)} :MQ:
\label{BRST-op}
\ee
on monomials $M$ and extend it by linearity to the whole ${\cal W}$. Then
$d_{Q}M \in {\cal W}$
and
\be
gh(d_{Q}M) = gh(M) -1.
\ee

The operator
$d_{Q}: {\cal W} \rightarrow {\cal W}$
is called the gauge variation; the properties of this
object are summarized in the following relations:
\be
d_{Q}^{2} = 0
\label{Q2}
\ee
\bea
d_{Q} h_{\mu\nu} = - {i\over 2}  
\left(\partial_{\mu}u_{\nu} + \partial_{\nu}u_{\mu} 
- \eta_{\mu\nu} \partial_{\rho} u^{\rho}
 \right) 
\nonumber \\
d_{Q} u_{\mu} = 0 \quad 
d_{Q} \tilde{u}_{\mu} = 
i\left( \partial^{\nu} h_{\mu\nu} + m v_{\mu} \right)
\nonumber \\
d_{Q} v_{\mu} = - {i m \over 2} u_{\mu}(x). 
\label{BRST}
\eea
\be
d_{Q}(MN) = (d_{Q}M) N + (-1)^{gh(M)} M (d_{Q}N), \quad
\forall M, N \in {\cal W}.
\label{Leibnitz}
\ee

If
$O: {\cal H}^{gh} \rightarrow {\cal H}^{gh}$
verifies the condition
\be
d_{Q} O = 0
\label{dQ}
\ee
then it induces a well defined operator
$[O]$
on the factor space
$\overline{Ker(Q)/Im(Q)}$.

Moreover, in this case the following formula is true for the matrix elements of
the factorized operator
$[O]$:
\be
([\Psi], [O] [\Phi]) = (\Psi, O \Phi).
\label{matrix-elem}
\ee

This kind of observables on the physical space will also be called {\it gauge
invariant observables}. An operator
$O: {\cal H}^{gh} \rightarrow {\cal H}^{gh}$
induces a gauge invariant observables if and only if it verifies:
\be
\left. d_{Q} O \right|_{Ker(Q)} = 0.
\ee
\label{gi-ghosts}

Not all operators verifying the condition (\ref{dQ}) are interesting. In fact,
the operators of the type
$d_{Q} O$
are inducing a null operator on the factor space; explicitly we have:
\be
[d_{Q} O] = 0.
\ee

In the framework of perturbative quantum field theory the axiom of 
factorization in the adiabatic limit is: XX
\be
\lim_{\epsilon \searrow 0} d_{Q} \int_{\R^{4}} dx 
\left. T_{n}(x_{1},\dots,x_{n}) g(\epsilon x) \right|_{Ker(Q)} = 0,
\quad \forall n \in \N^{*}.
\label{factorization1}
\ee

If infrared divergences cannot be avoided, the one can consider the preceding
relation at the heuristic level and impose the postulate:
\be
d_{Q} T_{n}(x_{1},\dots,x_{n}) = i \sum_{l=1}^{n}
{\partial \over \partial x^{\mu}_{l}} 
T^{\mu}_{n/l}(x_{1},\dots,x_{n}),
\quad \forall n \in \N^{*}
\label{gauge-n}
\ee
as it is done in \cite{Sc1}. In particular we have for
$n = 1$
\be
d_{Q} T_{1}(x) = i {\partial \over \partial x^{\mu}} T^{\mu}(x)
\label{gauge1}
\ee
for some Wick polynomials
$T^{\mu}(x)$.
The derivation of the most general expression of
$T_{1}$
can be done in the original variables
$H_{\mu\nu}, \Phi, u_{\mu}, \tilde{u}_{\mu}, v_{\mu}$
or with
$h_{\mu\nu}^{(\alpha)}$
for any values of $\alpha$. If we change the fields, we must correspondingly
change the expression of the operator
$d_{Q}$
and the final result should be the same. In formulae:
\be
d_{Q} T_{1}(H_{\mu\nu},\dots) 
= i {\partial \over \partial x^{\mu}} T^{\mu}(H_{\mu\nu},\dots) 
\quad \Leftrightarrow \quad
d_{Q} T_{1}(h^{(\alpha)}_{\mu\nu},\dots) 
= i {\partial \over \partial x^{\mu}} T^{\mu}(h^{(\alpha)}_{\mu\nu},\dots) 
\ee
for any $\alpha$.

\newpage

\section{First Order Gauge Invariance\label{first}}

In view of the discussion from the preceding Section it is natural to discard
from the interaction Lagrangian (the first-order chronological product $T_1$)
expressions of the type
\be
d_{Q}B + \partial_{\mu}B^{\mu}, \quad gh(B) = -1, gh(B^{\mu}) = 0;
\ee
we call such an expression a {\it trivial coupling}. If the difference
of two couplings is a trivial one then we call them {\it equivalent}.
In this Section we prove the following
\begin{thm}
Let us consider the most general Wick polynomial $T$ tri-linear in the fields
$
H_{\mu\nu}, \Phi, u_{\mu}, \tilde{u}_{\mu}, v_{\mu}
$
verifying the following conditions:
\bea
U_{g} T = T U_{g}, \quad \forall g \in {\cal P}
\nonumber \\
gh(T) = 0
\nonumber \\
3 \leq deg(T) \leq 5
\label{conditions}
\eea
and the gauge invariance condition (\ref{gauge1}). Then $T$ is equivalent to
the following expression:
\be
T = a~T^{(a)} + b~T^{(b)} \qquad a, b \in \R 
\ee
where
\bea
T^{(a)} \equiv [ 
- 2 H^{\mu\nu} (\partial_{\mu}H_{\rho\sigma}) (\partial_{\nu}H^{\rho\sigma})
- 4 H^{\mu\nu} (\partial^{\alpha}H^{\rho\mu}) (\partial^{\rho}H_{\alpha\nu})
+ 2 \Phi (\partial^{\mu}H^{\rho\sigma}) (\partial_{\rho}H_{\mu\sigma})
\nonumber \\
+ 4 \Phi H^{\rho\sigma} (\partial_{\rho}\partial^{\nu}H_{\nu\sigma})
+ \Phi^{2} (\partial_{\mu}\partial_{\nu}H^{\mu\nu})
+ (\partial^{\mu}\Phi) (\partial^{\nu}\Phi) H_{\mu\nu}
+ 4 H^{\mu\nu} u^{\rho} (\partial_{\mu}\partial_{\nu}\tilde{u}_{\rho})
\nonumber \\
+ 4 
(\partial_{\mu}H^{\mu\nu}) u^{\rho} (\partial_{\nu}\tilde{u}_{\rho})
+ 4 (\partial_{\mu}H^{\mu\nu}) u^{\rho} (\partial_{\rho}\tilde{u}_{\nu})
+ 2 (\partial^{\mu}\Phi) u^{\rho} (\partial_{\rho}\tilde{u}_{\mu})
+ 4 H^{\mu\nu} (\partial_{\mu}v_{\rho}) (\partial_{\nu}v^{\rho})
 ]
\nonumber \\
- 4 m (\partial_{\mu}v_{\nu}) u^{\mu} \tilde{u}^{\nu}
+ m^{2} \left( {2 \over 3} H^{\mu\nu} H_{\mu\rho} H_{\nu}^{\cdot\rho}
+ {1 \over 2}  H^{\mu\nu} H_{\mu\nu} \Phi
- {3 \over 4} \Phi^{3}
- \Phi u^{\mu} \tilde{u}_{\mu} 
+ \Phi v^{\mu} v_{\mu}
 \right)
\eea
and
\bea
T^{(b)} \equiv
- H^{\mu\nu} (\partial_{\mu}v_{\rho}) (\partial_{\nu}v^{\rho})
+ 2 H^{\mu\nu} (\partial_{\mu}v_{\nu}) (\partial_{\rho}v^{\rho})
- 2 H^{\mu\nu} (\partial_{\mu}v_{\rho}) (\partial^{\rho}v^{\nu})
\nonumber \\
+ m [ H^{\mu\nu} (\partial_{\rho}H_{\mu\nu}) v^{\rho}
- 2 H^{\mu\nu} (\partial_{\nu}H_{\mu\rho}) v^{\rho}
+ H^{\mu\nu} (\partial_{\mu}\Phi) v_{\nu}
+ (\partial_{\rho}v^{\rho}) u_{\mu} \tilde{u}^{\mu}
- {1 \over 2} (\partial_{\mu}v_{\nu}) u^{\mu} \tilde{u}^{\nu}
\nonumber \\
+ (\partial_{\nu}v_{\mu}) u^{\mu} \tilde{u}^{\nu}
+ v^{\nu} u^{\mu} (\partial_{\nu}\tilde{u}_{\mu})
+ v_{\mu} u^{\mu} (\partial^{\nu}\tilde{u}_{\mu})
- {1 \over 2} v^{\nu} u^{\mu} (\partial_{\mu}\tilde{u}_{\nu})
+ 2 v_{\mu} v_{\nu} (\partial^{\mu}v^{\nu}) ]
\nonumber \\
+ m^{2} \left( - {1 \over 3} H^{\mu\nu} H_{\mu\rho} H_{\nu}^{\cdot\rho}
- H^{\mu\nu} u_{\mu} \tilde
{u}_{\nu}
+ {3 \over 2} H^{\mu\nu} v_{\mu} v_{\nu}
 \right)
.
\eea

The preceding expression
 has a smooth limit
$m \rightarrow 0$. 
\end{thm}
{\bf Proof:}
We make a list of all Wick monomials verifying the conditions 
(\ref{conditions}). The condition of Lorentz invariance depends essentially 
on the dimension of the space-time. In other dimensions than $4$ the list
below changes drastically. Also the fact that
$H_{\mu\nu}$
is traceless is useful in eliminating many terms.
First we have terms which do also appear for the massless case i.e. without the 
ghost field
$
v_{\mu}
$
namely:
\newpage
\bea
T^{(1)} = c^{(1)} H^{\mu\nu} H_{\mu\rho} H_{\nu}^{\cdot\rho}
\nonumber \\
T^{(2)} = c^{(2)} H^{\mu\nu} H_{\mu\nu} \Phi
\nonumber \\
T^{(3)} = c^{(3)} \Phi^{3}
\nonumber \\
T^{(4)} = c^{(4)} H^{\mu\nu} u_{\mu} \tilde{u}_{\nu}
\nonumber \\
T^{(5)} = c^{(5)} \Phi u^{\mu} \tilde{u}_{\mu}
\nonumber \\
T^{(6)} = c^{(6)}_{1} H^{\mu\nu} (\partial_{\mu}H_{\rho\sigma})
(\partial_{\nu}H^{\rho\sigma})
+ c^{(6)}_{2} H^{\mu\nu} (\partial_{\rho}H_{\rho\sigma}) 
(\partial_{\sigma}H^{\mu\nu})
+ c^{(6)}_{3} H^{\mu\nu} (\partial_{\mu}H^{\rho\sigma})
(\partial_{\rho}H_{\nu\sigma})
\nonumber \\
+ c^{(6)}_{4} H^{\mu\nu} (\partial_{\mu}H_{\rho\nu})
(\partial_{\sigma}H^{\sigma\rho})
+ c^{(6)}_{5} H^{\mu\nu} (\partial^{\rho}H_{\rho\mu})
(\partial^{\sigma}H_{\sigma\nu})
+ c^{(6)}_{6} H^{\mu\nu} (\partial^{\alpha}H^{\rho\mu})
(\partial^{\rho}H_{\alpha\nu})
\nonumber \\
+ c^{(6)}_{7} \epsilon_{\mu\rho\alpha\lambda}
H^{\mu\nu} (\partial^{\lambda}H^{\rho}_{\cdot\nu}) 
(\partial_{\beta}H^{\alpha\beta})
+ c^{(6)}_{8} \epsilon_{\mu\rho\alpha\lambda}
H^{\mu\nu} (\partial^{\lambda}H^{\rho\sigma}) 
(\partial_{\sigma}H^{\alpha}_{\cdot\nu})
\nonumber \\
+ c^{(6)}_{9} \epsilon_{\mu\rho\alpha\lambda}
H^{\mu\nu} (\partial^{\lambda}H^{\rho\sigma}) 
(\partial_{\nu}H^{\alpha}_{\cdot\sigma})
\nonumber \\
T^{(7)} = c^{(7)}_{1} \Phi (\partial_{\mu}H^{\mu\sigma})
(\partial^{\nu}H_{\nu\sigma})
+ c^{(7)}_{2} \Phi (\partial^{\mu}H^{\rho\sigma}) (\partial_{\rho}H_{\mu\sigma})
+ c^{(7)}_{3} \epsilon_{\mu\nu\rho\alpha}
\Phi (\partial^{\mu}H^{\rho\sigma}) (\partial^{\nu}H^{\alpha}_{\cdot\sigma})
\nonumber \\
T^{(8)} = c^{(8)} \Phi H^{\rho\sigma} 
(\partial_{\rho}\partial^{\nu}H_{\nu\sigma})
\nonumber \\
T^{(9)} = c^{(9)} \Phi^{2} (\partial_{\mu}\partial_{\nu}H^{\mu\nu})
\nonumber \\
T^{(10)} = c^{(10)} (\partial^{\mu}\Phi) (\partial^{\nu}\Phi) H_{\mu\nu}
\nonumber \\
T^{(11)} = c^{(11)}_{1} (\partial_{\mu}H^{\mu\nu}) u_{\nu}
(\partial_{\rho}\tilde{u}^{\rho})
+ c^{(11)}_{2} (\partial_{\mu}H^{\mu\nu}) u^{\rho} 
(\partial_{\nu}\tilde{u}_{\rho})
+ c^{(11)}_{3} (\partial_{\mu}H^{\mu\nu}) u^{\rho}
(\partial_{\rho}\tilde{u}_{\nu})
\nonumber \\
+ c^{(11)}_{4} (\partial^{\rho}H^{\mu\nu}) u_{\mu}
(\partial_{\nu}\tilde{u}_{\rho})
+ c^{(11)}_{5} (\partial^{\rho}H^{\mu\nu}) u_{\rho}
(\partial_{\mu}\tilde{u}_{\nu})
\nonumber \\
+ c^{(11)}_{6} \epsilon_{\mu\rho\alpha\sigma}
(\partial^{\rho}H^{\mu\nu}) u^{\alpha} (\partial^{\sigma}\tilde{u}_{\nu})
+ c^{(11)}_{7} \epsilon_{\mu\rho\alpha\beta}
(\partial^{\rho}H^{\mu\nu}) u^{\alpha} (\partial_{\nu}\tilde{u}^{\beta})
+ c^{(11)}_{8} \epsilon_{\mu\sigma\alpha\beta}
(\partial_{\nu}H^{\mu\nu}) u^{\alpha} (\partial^{\sigma}\tilde{u}^{\beta})
\nonumber \\
T^{(12)} = c^{(12)}_{1} (\partial^{\rho}\partial_{\nu}H^{\mu\nu}) u_{\mu}
\tilde{u}_{\rho}
+ c^{(12)}_{2} (\partial^{\rho}\partial_{\nu}H^{\mu\nu}) u_{\rho} 
\tilde{u}_{\mu}
+ c^{(12)}_{3} (\partial_{\mu}\partial_{\nu}H^{\mu\nu}) u^{\rho} 
\tilde{u}_{\rho}
\nonumber \\
+ c^{(12)}_{4} \epsilon_{\mu\rho\alpha\beta}
(\partial^{\rho}\partial_{\nu}H^{\mu\nu}) u^{\alpha} \tilde{u}^{\beta}
\nonumber \\
T^{(13)} = c^{(13)}_{1} H^{\mu\nu} u_{\mu} 
(\partial_{\nu}\partial_{\rho}\tilde{u}_{\rho})
+ c^{(13)}_{2} H^{\mu\nu} u_{\rho} 
(\partial^{\rho}\partial_{\nu}\tilde{u}_{\mu})
+ c^{(13)}_{3} H^{\mu\nu} u^{\rho} 
(\partial_{\mu}\partial_{\nu}\tilde{u}_{\rho})
\nonumber \\
+ c^{(13)}_{4} \epsilon_{\mu\rho\alpha\beta} H^{\mu\nu} u^{\alpha} 
(\partial^{\rho}\partial_{\nu}\tilde{u}_{\beta})
\nonumber \\
T^{(14)} = c^{(14)}_{1} (\partial^{\mu}\Phi) u_{\mu} 
(\partial_{\rho}\tilde{u}^{\rho})
+ c^{(14)}_{2} (\partial^{\mu}\Phi) u^{\rho} 
(\partial_{\rho}\tilde{u}_{\mu})
+ c^{(14)}_{3} \epsilon_{\mu\nu\alpha\beta} (\partial^{\mu}\Phi) u^{\alpha} 
(\partial^{\nu}\tilde{u}^{\beta})
\nonumber \\
T^{(15)} = c^{(15)} \Phi u_{\mu} (\partial^{\mu}\partial^{\nu}\tilde{u}_{\nu})
\nonumber \\
T^{(16)} = c^{(15)} (\partial^{\mu}\partial^{\nu}\Phi) u_{\mu} \tilde{u}_{\nu};
\eea
then we have the terms containing at least one factor
$
v_{\mu}
$
namely:
\bea
U^{(1)} = d^{(1)} H^{\mu\nu} v_{\mu} v_{\nu}
\nonumber \\
U^{(2)} = d^{(2)} \Phi v^{\mu} v_{\mu} 
\nonumber \\
U^{(3)} = d^{(3)}_{1} H^{\mu\nu} (\partial_{\alpha}H_{\mu\nu}) A^{\alpha}
+ d^{(3)}_{2} H^{\mu\nu} (\partial^{\alpha}H_{\mu\alpha}) v_{\nu}
\nonumber \\
+ d^{(3)}_{3} H^{\mu\nu} (\partial_{\nu}H_{\mu\alpha}) v^{\alpha}
+ d^{(3)}_{4} \epsilon_{\mu\rho\alpha\beta} H^{\mu\nu} 
(\partial^{\alpha}H^{\rho}_{\cdot\nu}) v^{\beta}
\nonumber \\
U^{(4)} = d^{(4)} H^{\mu\nu} (\partial_{\mu}\Phi) v_{\nu} 
\nonumber \\
U^{(5)} = d^{(5)} \Phi (\partial^{\nu}H_{\mu\nu}) v_{\mu} 
\nonumber \\
U^{(6)} = d^{(6)} \Phi (\partial^{\alpha}\Phi) v_{\alpha} 
\nonumber \\
U^{(7)} = d^{(7)}_{1} (\partial_{\alpha}v^{\alpha}) u_{\mu} \tilde{u}^{\mu}
+ d^{(7)}_{2} (\partial_{\mu}v_{\nu}) u^{\mu} \tilde{u}^{\nu}
+ d^{(7)}_{3} (\partial_{\nu}v_{\mu}) u^{\mu} \tilde{u}^{\nu}
+ d^{(7)}_{4} \epsilon_{\mu\nu\alpha\beta}
(\partial^{\alpha}v^{\beta}) u^{\mu} \tilde{u}^{\nu}
\nonumber \\
U^{(8)} = d^{(8)}_{1} v^{\alpha} u^{\mu} (\partial_{\alpha}\tilde{u}_{\mu})
+ d^{(8)}_{2} v_{\mu} u^{\mu} (\partial_{\nu}\tilde{u}^{\nu})
+ d^{(8)}_{3} v^{\nu} u^{\mu} (\partial_{\mu}\tilde{u}_{\nu})
+ d^{(8)}_{4} \epsilon_{\mu\nu\alpha\beta}
v^{\alpha} u^{\mu} (\partial^{\beta}\tilde{u}^{\nu})
\nonumber \\
U^{(9)} = d^{(9)} v_{\mu} v_{\nu} (\partial^{\mu}v^{\nu}) 
\nonumber \\
U^{(10)} = d^{(10)}_{1} H^{\mu\nu} (\partial_{\mu}v_{\alpha})
(\partial_{\nu}v^{\alpha})
+ d^{(10)}_{2} H^{\mu\nu} (\partial_{\mu}v_{\nu}) (\partial_{\beta}v^{\beta})
+ d^{(10)}_{3} H^{\mu\nu} (\partial_{\mu}v_{\rho}) (\partial^{\rho}v_{\nu})
\nonumber \\
+ d^{(10)}_{4} \epsilon_{\mu\rho\alpha\beta}
H^{\mu\nu} (\partial^{\rho}v_{\nu}) (\partial^{\alpha}v^{\beta})
+ d^{(10)}_{5} \epsilon_{\mu\rho\alpha\beta}
H^{\mu\nu} (\partial^{\rho}v^{\alpha}) (\partial_{\nu}v^{\beta})
\nonumber \\
U^{(11)} = d^{(11)}_{1} H^{\mu\nu} v^{\alpha}
(\partial_{\mu}\partial_{\nu}v_{\alpha})
+ d^{(11)}_{2} H^{\mu\nu} v_{\nu} (\partial_{\mu}\partial_{\beta}v^{\beta})
+ d^{(11)}_{3} H^{\mu\nu} v^{\rho} (\partial_{\mu}\partial_{\rho}v_{\nu})
\nonumber \\
+ d^{(11)}_{4} \epsilon_{\mu\alpha\rho\beta}
H^{\mu\nu} v^{\alpha} (\partial^{\rho}\partial_{\nu}v^{\beta})
\nonumber \\
U^{(12)} = d^{(12)}_{1} \Phi (\partial_{\alpha}v^{\alpha})^{2}
+ d^{(12)}_{2} \Phi (\partial^{\mu}v^{\alpha}) (\partial_{\alpha}v_{\mu})
+ d^{(12)}_{3} \epsilon_{\mu\nu\alpha\beta} \Phi
(\partial^{\mu}v^{\alpha}) (\partial^{\nu}v^{\beta})
\nonumber \\
U^{(13)} = d^{(13)} \Phi v^{\alpha} (\partial_{\alpha}\partial_{\beta}v^{\beta})
\nonumber \\
U^{(14)} = d^{(14)}_{1} v^{\alpha} (\partial^{\beta}v_{\beta})
(\partial_{\alpha}\partial_{\nu}v^{\nu})
+ d^{(14)}_{2} v^{\alpha} (\partial_{\alpha}v_{\beta})
(\partial^{\beta}\partial^{\nu}v_{\nu})
\nonumber \\
+ d^{(14)}_{3} \epsilon_{\mu\nu\alpha\beta} 
v^{\alpha} (\partial^{\mu}v_{\beta}) (\partial^{\nu}\partial^{\rho}v_{\rho})
+ d^{(14)}_{4} \epsilon_{\mu\nu\alpha\beta} 
v_{\rho} (\partial^{\mu}v^{\beta}) (\partial^{\nu}\partial^{\rho}v^{\alpha})
\nonumber \\
+ d^{(14)}_{5} \epsilon_{\mu\nu\alpha\beta} 
v^{\alpha} (\partial^{\mu}v^{\rho}) (\partial^{\nu}\partial_{\rho}v^{\beta})
\nonumber \\
U^{(15)} = d^{(15)} v^{\alpha} v^{\beta} 
(\partial_{\alpha}\partial_{\beta}\partial_{\mu}v^{\mu}).
\eea

We have discarded a lot of terms because up to a total derivatives they are
of the type already considered. As a general strategy, we have eliminated all 
terms with derivatives on the 
ghost fields
$u_{\mu}$.
It is somewhat more complicated to prove that one can make
$c^{(6)}_{3} = 0$
if one subtracts a total divergence and redefines  
$c^{(6)}_{4}, c^{(6)}_{5}, c^{(6)}_{6}$,
and 
$c^{(6)}_{9} = 0$
if one subtract a total divergence and redefines  
$c^{(6)}_{7}, c^{(6)}_{8}$.

Also because
\bea
(\partial^{2} + m^{2}_{j}) f_{j} = 0, \quad j = 1,2,3 
\quad \Longrightarrow
\nonumber \\
(\partial^{\mu}f_{1}) (\partial_{\mu}f_{2}) f_{3}
= {1\over 2} (m^{2}_{1} + m^{2}_{2} - m^{2}_{3})~f_{1} f_{2} f_{3}
+ {1\over 2} \partial_{\mu} \Bigl[ (\partial^{\mu}f_{1}) f_{2} f_{3}
+ f_{1} (\partial^{\mu}f_{2}) f_{3} - 
f_{1} f_{2} (\partial^{\mu}f_{3}) \Bigl]
\label{magic}
\eea
we can eliminate many terms by subtracting a total divergence and redefining
other terms of lower canonical dimension. 

Now we can put to zero some of the constants above if we subtract from $T$
a coboundary i.e. an expression of the form
$d_{Q} B$
where we take $B$ to be a Wick polynomial with the following properties: 
\bea
U_{g} B = B U_{g}, \quad \forall g \in {\cal P}
\nonumber \\
gh(T) = - 1
\nonumber \\
2 \leq deg(T) \leq 4.
\eea

We have the following admissible expressions:

First we have terms which do also appear for the massless case i.e. without the 
ghost field
$
v_{\mu}
$
namely:
\bea
B^{(1)} = b^{(1)}_{1} H^{\mu\nu} (\partial_{\rho}H_{\mu\nu}) \tilde{u}^{\rho}
+ b^{(1)}_{2} H^{\mu\nu} (\partial_{\mu}H_{\mu\rho}) \tilde{u}^{\rho}
\nonumber \\
+ b^{(1)}_{3} H^{\mu\nu} (\partial^{\rho}H_{\mu\rho}) \tilde{u}_{\nu}
+ b^{(1)}_{4} \epsilon_{\mu\rho\alpha\beta}
H^{\mu\nu} (\partial^{\alpha}{H^{\rho}}_{\cdot\nu}) \tilde{u}^{\beta}
\nonumber \\
B^{(2)} = b^{(2)} \Phi (\partial_{\mu}H^{\mu\nu}) \tilde{u}_{\nu} 
\nonumber \\
B^{(3)} = b^{(3)} \Phi H^{\mu\nu} (\partial_{\mu}\tilde{u}_{\nu}) 
\nonumber \\
B^{(4)} = b^{(4)} \Phi^{2} \partial^{\mu}\tilde{u}_{\mu}
\nonumber \\
B^{(5)} = b^{(5)}_{1} u^{\mu} \tilde{u}_{\mu} (\partial_{\rho}\tilde{u}^{\rho})
+ b^{(5)}_{2} u^{\mu} \tilde{u}^{\nu} (\partial_{\mu}\tilde{u}_{\nu})
+ b^{(5)}_{3} u^{\mu} \tilde{u}^{\nu} (\partial_{\nu}\tilde{u}_{\mu})
+ b^{(5)}_{4} \epsilon_{\mu\nu\rho\sigma} u^{\mu} \tilde{u}^{\nu} 
(\partial^{\rho}\tilde{u}^{\sigma});
\eea
then we have the terms containing at least one factor
$
v_{\mu}
$
namely:
\bea
V^{(1)} = f^{(1)} H^{\mu\nu} v_{\mu} \tilde{u}_{\nu}
\nonumber \\
V^{(2)} = f^{(2)} \Phi v^{\mu} \tilde{u}_{\mu} 
\nonumber \\
V^{(3)} = f^{(3)}_{1} v^{\alpha} (\partial_{\mu}v_{\alpha}) \tilde{u}^{\mu}
+ f^{(3)}_{2} v_{\mu} (\partial^{\alpha}v_{\alpha}) \tilde{u}^{\mu}
+ f^{(3)}_{3} v_{\nu} (\partial^{\nu}v^{\mu}) \tilde{u}_{\mu}
+ f^{(3)}_{4} \epsilon_{\mu\nu\alpha\beta}
v^{\alpha} (\partial^{\nu}v^{\beta}) \tilde{u}^{\mu}.
\eea

We have discarded some terms because up to a total derivatives they are
of the type already considered. Now one can prove that:
\begin{itemize}
\item
One can use
$B^{(1)}$
to make
$c^{(6)}_{2}, c^{(6)}_{4}, c^{(6)}_{5}, c^{(6)}_{7}$
equal to zero. In this way one redefines
$T^{(7)}, T^{(8)}, T^{(11)}, T^{(12)}, T^{(13)}$.
\item
One can use
$B^{(2)}$
to make
$c^{(7)}_{1}$
equal to zero. In this way one redefines
$T^{(9)}, T^{(11)}, T^{(12)}, \\ T^{(14)}, T^{(15)}, T^{(16)}$.
\item
One can use
$B^{(3)}$
to make
$c^{(13)}_{2}$
equal to zero. In this way one redefines
$T^{(8)} - T^{(11)}, \\ T^{(14)}, T^{(15)}$.
\item
One can use
$B^{(4)}$
to make
$c^{(15)}$
equal to zero. In this way one redefines
$T^{(9)}, T^{(14)}$.
\item
One can use
$B^{(5)}$
to make
$c^{(12)}_{1}, c^{(12)}_{2}, c^{(12)}_{3}, c^{(12)}_{4}$
equal to zero. In this way one redefines
$T^{(11)}, T^{(14)}, T^{(16)}$.
\item
One can use
$V^{(1)}$
to make
$d^{(3)}_{2}$
equal to zero. In this way one redefines
$T^{(4)}, U^{(1)}, U^{(4)}, \\ U^{(7)}, U^{(8)}$.
\item
One can use
$V^{(2)}$
to make
$d^{(5)}$
equal to zero. In this way one redefines
$T^{(5)}, U^{(2)}, U^{(6)}, \\ U^{(7)}, U^{(8)}$.
\item
One can use 
$V^{(3)}$
to make XX
$d^{(7)}_{1}, d^{(7)}_{2}, d^{(7)}_{3}, d^{(7)}_{4}$
equal to zero. In this way one redefines
$U^{(7)} - U^{(10)}, U^{(12)}, U^{(13)}$.
\end{itemize} 

One can count that we are left with 37 coefficients of type 
$
c^{(j)}
$
and 23 coefficients of type
$
d^{(j)}.
$
So we have 60 free parameters which should be fixed by the condition of
gauge invariance. 

Now one considers the coupling
\be
T = \sum_{j=1}^{16} T^{(j)} +  \sum_{j=1}^{13} U^{(j)} 
\ee
and computes the expression
$d_{Q}T$.
It is convenient to follow the same strategy as above and eliminate, up to a 
divergence, the derivatives on the ghost fields
$u_{\mu}$.
One gets by a straightforward but tedious computation the following result:
\be
d_{Q}T = i~\partial_{\mu}X^{\mu} + i~u^{\mu}~Y_{\mu} + i~Z_{4} + i~Z_{6}
\label{dQT}
\ee
where the expressions
$X^{\mu}, Y^{\mu}$
do not contain ghost fields and the expressions
$Z_{j}$
are tri-linear in the ghost fields of canonical dimension 
$j = 4,6$.
The explicit expression of
$X^{\mu}$
and
$Z_{j}$
are not important for the moment. The expression of 
$Y^{\mu}$
is rather long: we group the various sectors:
\be
Y_{\mu} = \sum_{j=1}^{22} Y_{\mu}^{j}
\label{y}
\ee
where
\bea
Y_{\mu}^{1} \equiv 
\left[ 3 c^{(1)} + m^{2} c^{(6)}_{6} - {1\over 2} m d^{(3)}_{3} \right]
H^{\nu\rho} (\partial_{\nu}H_{\mu\rho})
+ \left[ 3 c^{(1)} - c^{(4)} + m^{2} c^{(6)}_{6} \right]
H_{\mu\nu} (\partial_{\rho}H^{\nu\rho})
\nonumber \\
- \left[ {3\over 2} c^{(1)} + c^{(2)} + m^{2} c^{(6)}_{1} 
+ {1\over 2} m d^{(3)}_{1} \right]
H^{\nu\rho} (\partial_{\mu}H_{\nu\rho})
+ {1\over 2} m d^{(3)}_{4} \epsilon_{\mu\nu\rho\sigma} H^{\sigma\lambda} 
(\partial^{\rho}H^{\nu}_{\cdot\lambda})
\label{y1}
\eea
\be
Y_{\mu}^{2} \equiv  
\left[ 2 c^{(2)} - c^{(5)} + m^{2} c^{(7)}_{2} 
- m^{2} c^{(8)} + {1\over 2} m^{2} c^{(11)}_{1} \right]
\Phi (\partial^{\nu}H_{\mu\nu})
\label{y2}
\ee
\be
Y_{\mu}^{3} \equiv  
\left[ 2 c^{(2)} - {1\over 2} c^{(4)} + m^{2} c^{(7)}_{2}
- {1\over 2} m^{2} c^{(8)} + {1\over 2} m^{2} c^{(13)}_{1} 
- {1\over 2} m d^{(4)} \right]
H^{\mu\nu} (\partial_{\nu}\Phi)
\label{y3}
\ee
\be
Y_{\mu}^{4} \equiv  
- \left[ 3 c^{(3)} + {1\over 2} c^{(5)} + {3\over 2} m^{2} c^{(9)}
+ m^{2} c^{(10)} - {1\over 2} m^{2} c^{(14)}_{1} 
+ {1\over 2} m d^{(6)} \right]
\Phi (\partial_{\mu}\Phi)
\label{y4}
\ee
\bea
Y_{\mu}^{5} \equiv  
{1\over 2} c^{(6)}_{1}
(\partial_{\mu}\partial_{\nu}H_{\rho\sigma}) (\partial^{\nu}H^{\rho\sigma})
+ \left[ - 2 c^{(6)}_{1} + c^{(6)}_{6} - c^{(11)}_{4} \right]
(\partial_{\nu}\partial^{\rho}H^{\nu\sigma}) (\partial_{\sigma}H_{\mu\rho})
\nonumber \\
- \left[ 2 c^{(6)}_{1} + c^{(11)}_{2} \right]
(\partial_{\sigma}\partial^{\rho}H_{\mu\rho}) (\partial_{\nu}H^{\nu\sigma})
+ \left[ - 2 c^{(6)}_{1} + c^{(6)}_{6} \right]
(\partial_{\sigma}\partial_{\rho}H_{\mu\rho}) (\partial^{\rho}H^{\nu\sigma})
\nonumber \\
- \left[ {1\over 2} c^{(6)}_{6} + c^{(7)}_{2} \right]
(\partial_{\mu}\partial_{\rho}H_{\nu\sigma}) (\partial^{\sigma}H^{\nu\rho})
- c^{(6)}_{6}
(\partial_{\rho}\partial_{\sigma}H_{\mu\nu}) (\partial^{\sigma}H^{\nu\rho})
- c^{(6)}_{6}
(\partial_{\rho}\partial_{\sigma}H^{\nu\rho}) (\partial^{\sigma}H_{\mu\nu})
\nonumber \\
- \left[ {1 \over 2} c^{(6)}_{6} + {1\over 2} c^{(8)} + c^{(11)}_{5} \right]
(\partial^{\nu}\partial_{\nu}H^{\nu\sigma}) (\partial_{\mu}H^{\rho\sigma})
- \left[ c^{(6)}_{6} + c^{(11)}_{3} \right]
(\partial_{\mu}\partial^{\rho}H_{\rho\sigma}) (\partial_{\nu}H^{\nu\sigma})
\nonumber \\
+ 2 c^{(6)}_{8} \epsilon_{\nu\rho\sigma\lambda}
(\partial_{\nu}\partial_{\tau}H^{\sigma}_{\cdot\mu}) 
(\partial^{\lambda}H^{\rho\tau})
+ 2 c^{(6)}_{8} \epsilon_{\mu\rho\sigma\lambda}
(\partial_{\nu}\partial^{\lambda}H^{\rho\tau}) (\partial_{\tau}H^{\nu\sigma})
\nonumber \\
+ \left[ 2 c^{(6)}_{8} + c^{(11)}_{7} \right] 
\epsilon_{\mu\alpha\beta\rho}
(\partial_{\nu}\partial_{\lambda}H^{\lambda\beta}) 
(\partial^{\rho}H^{\nu\alpha})
+ \left[ 2 c^{(6)}_{8} - c^{(7)}_{3} \right] 
\epsilon_{\alpha\beta\rho\lambda}
(\partial_{\mu}\partial^{\lambda}H^{\nu\beta}) 
(\partial^{\rho}H^{\alpha}_{\cdot\nu})
\nonumber \\
+ \left[ c^{(6)}_{8} - c^{(11)}_{3} \right] 
\epsilon_{\mu\rho\alpha\lambda}
(\partial_{\sigma}\partial^{\lambda}H^{\rho\sigma}) 
(\partial_{\nu}H^{\alpha\nu})
- c^{(11)}_{1}
(\partial^{\rho}\partial^{\sigma}H_{\rho\sigma}) (\partial^{\nu}H_{\mu\nu})
\nonumber \\
- c^{(11)}_{6}  
\epsilon_{\mu\sigma\alpha\rho}
(\partial^{\sigma}\partial^{\lambda}H_{\nu\lambda}) 
(\partial^{\rho}H^{\alpha\nu}) \qquad
\label{y5}
\eea
\bea
Y_{\mu}^{6} \equiv  
- \left[ 2 c^{(6)}_{1} + c^{(13)}_{3} \right]
H^{\nu\sigma} (\partial_{\nu}\partial_{\sigma}\partial^{\rho}H_{\mu\rho}) 
- {1\over 2} \left[ c^{(6)}_{1} + c^{(8)} \right]
H^{\nu\sigma} (\partial_{\mu}\partial_{\nu}\partial^{\rho}H_{\rho\sigma}) 
\nonumber \\
+ \left[ c^{(6)}_{8} - c^{(13)}_{4} \right] 
\epsilon_{\mu\rho\alpha\lambda}
H^{\nu\alpha} (\partial_{\sigma}\partial_{\nu}\partial^{\lambda}H^{\rho\sigma}) 
- c^{(13)}_{1}
H_{\mu\nu} (\partial^{\nu}\partial^{\rho}\partial^{\sigma}H_{\rho\sigma}) 
\label{y6}
\eea
\bea
Y_{\mu}^{7} \equiv  
- \left[ c^{(7)}_{2} + {1\over 2} c^{(11)}_{4} \right]
(\partial^{\rho}\partial^{\sigma}\Phi) (\partial_{\sigma}H_{\mu\rho})
+ \left[ - c^{(7)}_{2} + {1\over 4} c^{(8)}
+ c^{(10)} - {1\over 2} c^{(11)}_{5} \right]
(\partial^{\sigma}\partial^{\rho}\Phi) (\partial_{\mu}H_{\rho\sigma})
\nonumber \\
+ \left[ {1\over 2} c^{(7)}_{2} + {1\over 2} c^{(8)}
+ c^{(10)} - {1\over 2} c^{(11)}_{2} 
- {1\over 2} c^{(11)}_{3} - c^{(16)} \right]
(\partial_{\mu}\partial_{\nu}\Phi) (\partial_{\rho}H^{\nu\rho})
\nonumber \\
+ \left[ c^{(7)}_{3} - {1\over 2} c^{(11)}_{6} - {1\over 2} c^{(11)}_{7}\right] 
\epsilon_{\mu\nu\rho\beta}
(\partial^{\nu}\partial_{\sigma}\Phi) (\partial^{\beta}H^{\rho\sigma}) \quad
\label{y7}
\eea
\bea
Y_{\mu}^{8} \equiv  
\left[ - c^{(7)}_{2} + {1\over 2} c^{(8)} \right]
(\partial_{\sigma}\Phi) (\partial^{\rho}\partial^{\sigma}H_{\mu\rho})
+ \left[ - {3 \over 2} c^{(7)}_{2} + c^{(8)} + c^{(10)} - c^{(14)}_{2} \right]
(\partial_{\rho}\Phi) (\partial_{\sigma}\partial_{\mu}H^{\rho\sigma})
\nonumber \\
+ \left[ {1\over 2} c^{(7)}_{2} - c^{(9)}  - c^{(14)}_{1} \right]
(\partial_{\mu}\Phi) (\partial_{\nu}\partial_{\rho}H^{\nu\rho})
+ \left[ c^{(7)}_{3} + c^{(13)}_{4} \right] 
\epsilon_{\mu\nu\rho\beta}
(\partial^{\nu}\Phi) (\partial_{\sigma}\partial^{\beta}H^{\rho\sigma})
\label{y8}
\eea
\be
Y_{\mu}^{9} \equiv  
\left[ - {1 \over 2} c^{(7)}_{2} + {1\over 2} c^{(8)}  - c^{(9)} \right]
\Phi (\partial_{\mu}\partial_{\nu}\partial_{\rho}H^{\nu\rho})
\label{y9}
\ee
\be
Y_{\mu}^{10} \equiv  
\left[ {1 \over 4} c^{(8)} + c^{(10)}  - {1 \over 2} c^{(13)}_{3} \right]
H^{\rho\sigma}  (\partial_{\mu}\partial_{\rho}\partial_{\sigma}\Phi)
\label{y10}
\ee
\be
Y_{\mu}^{11} \equiv  
{1 \over 2} \left[ c^{(10)} - c^{(14)}_{2} - c^{(16)} \right]
(\partial^{\nu}\Phi) (\partial_{\mu}\partial_{\nu}\Phi)
\label{y11}
\ee
\be
Y_{\mu}^{12} \equiv  
- \left[ m c^{(4)} + m d^{(1)} + {1 \over 2} m^{2} d^{(3)}_{3} \right]
H_{\mu\nu} v^{\nu}
\label{y12}
\ee
\be
Y_{\mu}^{13} \equiv  
- \left[ m c^{(5)} + m d^{(2)} + {1 \over 2} m^{2} d^{(4)} 
- {1 \over 2} m^{2} d^{(8)}_{2}  \right]
\Phi  v_{\mu}
\label{y13}
\ee
\bea
Y_{\mu}^{14} \equiv  
- \left[ m c^{(11)}_{1} + d^{(3)}_{1} + {1 \over 2} d^{(3)}_{3} 
+ d^{(7)}_{1} - {1 \over 2} m d^{(10)}_{2} \right]
(\partial^{\nu}H_{\mu\nu}) (\partial_{\alpha}v^{\alpha})
\nonumber \\
- \left[ m c^{(11)}_{2} + d^{(3)}_{3} + d^{(7)}_{3} - m d^{(10)}_{1} \right]
(\partial_{\rho}H^{\nu\rho}) (\partial_{\nu}v_{\mu})
\nonumber \\
- \left[ m c^{(11)}_{3} - { 1\over 2} d^{(4)} + d^{(7)}_{2} 
- {1 \over 2} m d^{(10)}_{3} \right]
(\partial_{\rho}H^{\nu\rho}) (\partial_{\mu}v_{\nu})
\nonumber \\
- \left[ m c^{(11)}_{4} + { 1\over 2} d^{(3)}_{3} - {1 \over 2} m d^{(10)}_{3} 
\right]
(\partial_{\rho}H_{\mu\nu}) (\partial^{\nu}v^{\rho})
\nonumber \\
- \left[ m c^{(11)}_{5} - { 3\over 4} d^{(3)}_{3} - {1 \over 2} d^{(4)} 
- {1 \over 2} m d^{(10)}_{2} \right]
(\partial_{\mu}H_{\rho\sigma}) (\partial^{\rho}v^{\sigma})
\nonumber \\
+ {1\over 2} d^{(3)}_{3}
(\partial_{\nu}H_{\mu\rho}) (\partial^{\nu}A_{\rho})
+ \left[ c^{(11)}_{6} - {1\over 2} m d^{(10)}_{4} \right]
\epsilon_{\nu\rho\alpha\beta}
(\partial^{\rho}H^{\alpha\nu}) (\partial^{\beta}v_{\nu})
\nonumber \\
+ \left[ c^{(11)}_{7} - {1 \over 2} d^{(3)}_{4} 
- {1 \over 2} m d^{(10)}_{5}\right] 
\epsilon_{\mu\rho\alpha\beta}
(\partial^{\rho}H^{\alpha\nu}) (\partial_{\nu}v^{\beta})
\nonumber \\
- \left[ c^{(11)}_{8} - {1 \over 2} d^{(3)}_{4} 
+ d^{(7)}_{4} - {1 \over 2} m d^{(10)}_{5} \right] 
\epsilon_{\mu\alpha\sigma\beta}
(\partial_{\nu}H^{\alpha\nu}) (\partial^{\sigma}v^{\beta})
\nonumber \\
- {1 \over 2} \left[ 3 d^{(3)}_{4} + m d^{(10)}_{4} \right] 
\epsilon_{\alpha\rho\nu\beta}
(\partial^{\alpha}H^{\rho}_{\cdot\mu}) (\partial^{\nu}v^{\beta})
\label{y14}
\eea
\bea
Y_{\mu}^{15} \equiv  
- \left[ m c^{(13)}_{1} + d^{(3)}_{1} + {1 \over 2} d^{(3)}_{3} 
- { 1 \over 2} m d^{(10)}_{2} - {1 \over 2} m d^{(10)}_{3} \right]
H_{\mu\nu} (\partial^{\nu}\partial^{\alpha}v_{\alpha})
\nonumber \\
- \left[ m c^{(13)}_{3} + {1 \over 2} d^{(3)}_{3} - m d^{(10)}_{1} \right]
H^{\rho\sigma} (\partial_{\rho}\partial_{\sigma}v_{\mu})
\nonumber \\
+ {1 \over 2} \left[ { 1 \over 2} d^{(3)}_{3} + d^{(4)} + m d^{(10)}_{2} 
+ m d^{(10)}_{3} \right]
H^{\nu\rho} (\partial_{\mu}\partial_{\nu}v_{\rho})
\nonumber \\
- \left[ m c^{(13)}_{4} - {1\over 2} m d^{(3)}_{4} - m d^{(10)}_{5} \right]
\epsilon_{\mu\alpha\rho\beta}
H^{\alpha\nu} (\partial_{\nu}\partial^{\rho}v^{\beta})
\label{y15}
\eea
\bea
Y_{\mu}^{16} \equiv  
- \left[ m c^{(14)}_{1} - { 1 \over 2} d^{(4)} - {1 \over 2} d^{(6)} 
+ { 1 \over 2} d^{(7)}_{1} - m d^{(12)}_{1} + {1 \over 2} m d^{(13)} \right]
(\partial_{\mu}\Phi) (\partial^{\alpha}v_{\alpha})
\nonumber \\
- \left[ m c^{(14)}_{2} + {1 \over 4} d^{(4)} + {1 \over 2} d^{(7)}_{2} 
- m d^{(12)}_{2} + {1 \over 2} d^{(13)} \right]
(\partial^{\alpha}\Phi) (\partial_{\mu}v_{\alpha})
\nonumber \\
+ {1 \over 2} \left[ d^{(4)} - d^{(7)}_{3} \right]
(\partial^{\nu}\Phi) (\partial_{\nu}v_{\mu})
\nonumber \\
+ m \left[ c^{(14)}_{3} - {1\over 2} d^{(7)}_{4} + d^{(12)}_{3} \right]
\epsilon_{\mu\nu\alpha\beta}
(\partial^{\nu}\Phi) (\partial^{\alpha}v^{\beta})
\label{y16}
\eea
\be
Y_{\mu}^{17} \equiv  
- \left[ m c^{(16)} - { 1 \over 4} d^{(4)} + {1 \over 2} d^{(8)}_{1} 
+ { 1 \over 2} d^{(8)}_{3} + {1 \over 2} m d^{(13)} \right]
(\partial_{\mu}\partial_{\nu}\Phi) v^{\nu}
\label{y17}
\ee
\bea
Y_{\mu}^{18} \equiv  
\left[ d^{(1)} - m d^{(7)}_{1} - m d^{(8)}_{2} + {1 \over 2} m d^{(9)} 
- {1 \over 2} m^{2} d^{(10)}_{2} \right]
v_{\mu} (\partial^{\alpha}v_{\alpha})
\nonumber \\
+ \left[ d^{(1)} - m d^{(7)}_{3} - m d^{(8)}_{1} 
- {1 \over 2} m^{2} d^{(10)}_{3} \right]
v^{\alpha} (\partial_{\alpha}v_{\mu})
\nonumber \\
- \left[ {1 \over 2} d^{(1)} + d^{(2)} + m d^{(7)}_{2} + m d^{(8)}_{3}  
+ {1 \over 2} m d^{(9)} + m^{2} d^{(10)}_{1} \right]
v^{\alpha} (\partial_{\mu}v_{\alpha})
\nonumber \\
- m \left[ d^{(7)}_{4} + d^{(8)}_{4} + {1 \over 2} m d^{(10)}_{5} \right]
\epsilon_{\mu\nu\alpha\beta}
v^{\nu} (\partial^{\alpha}v^{\beta})
\label{y18}
\eea
\bea
Y_{\mu}^{19} \equiv  
\left[ {1 \over 2} d^{(3)}_{3} + {1 \over 2} d^{(4)} - d^{(8)}_{3} \right]
v^{\rho} (\partial_{\mu}\partial_{\nu}H_{\nu\rho})
- \left[ {1 \over 2} d^{(3)}_{3} + d^{(8)}_{1} \right]
v_{\rho} (\partial^{\nu}\partial^{\rho}H_{\mu\nu})
\nonumber \\
- \left[ {1 \over 2} d^{(3)}_{3} + d^{(8)}_{2} \right]
v_{\mu} (\partial_{\mu}\partial_{\rho}H^{\nu\rho})
- \left[ d^{(3)}_{4} + d^{(8)}_{4} \right]
\epsilon_{\mu\alpha\rho\beta}
v^{\beta} (\partial^{\alpha}\partial_{\nu}H^{\nu\rho})
\label{y19}
\eea
\be
Y_{\mu}^{20} \equiv  
\left[ {1 \over 2} d^{(6)} + m d^{(12)}_{1} + m d^{(12)}_{2} - m d^{(13)} 
\right]
\Phi (\partial_{\mu}\partial_{\nu}v^{\nu})
\label{y20}
\ee
\bea
Y_{\mu}^{21} \equiv  
{1 \over 2} d^{(10)}_{1} 
(\partial^{\nu}v^{\alpha}) (\partial_{\mu}\partial_{\nu}v_{\alpha})
+ {1 \over 2} \left[ d^{(10)}_{2} + d^{(10)}_{3} - d^{(13)} \right]
(\partial_{\mu}v_{\nu}) (\partial^{\nu}\partial^{\rho}v_{\rho})
\nonumber \\
+ {1 \over 2} d^{(10)}_{2}   
(\partial_{\nu}v_{\mu}) (\partial^{\nu}\partial^{\rho}v_{\rho})
+ {1 \over 2} d^{(10)}_{3}   
(\partial^{\nu}v^{\rho}) (\partial_{\nu}\partial_{\rho}v_{\mu})
\nonumber \\
- d^{(12)}_{1}   
(\partial_{\nu}v^{\nu}) (\partial_{\mu}\partial_{\rho}v^{\rho})
- d^{(12)}_{2}   
(\partial^{\beta}v^{\alpha}) (\partial_{\mu}\partial_{\alpha}v_{\beta})
\nonumber \\
+ {1 \over 2} d^{(10)}_{4} 
\epsilon_{\mu\rho\alpha\beta}
(\partial^{\alpha}v^{\beta}) (\partial^{\rho}\partial^{\nu}v_{\nu})
+ {1 \over 2} d^{(10)}_{4} 
\epsilon_{\mu\rho\alpha\beta}
(\partial^{\rho}v^{\nu}) (\partial_{\nu}\partial^{\alpha}v^{\beta})
\nonumber \\
+ \left[ {1 \over 2} d^{(10)}_{4} + d^{(12)}_{3} \right]
\epsilon_{\rho\nu\alpha\beta}
(\partial^{\alpha}v^{\beta}) (\partial_{\mu}\partial^{\rho}v^{\nu})
+ {1 \over 2} d^{(10)}_{5} 
\epsilon_{\mu\rho\alpha\beta}
(\partial_{\nu}v^{\beta}) (\partial^{\nu}\partial^{\rho}v^{\alpha})
\label{y21}
\eea
\be
Y_{\mu}^{22} \equiv  
- { 1 \over 2} d^{(13)}  
v^{\alpha} (\partial_{\mu}\partial_{\alpha}\partial_{\beta}v^{\beta}). 
\label{y22}
\ee

We now impose the gauge invariance condition (\ref{gauge1}). It is sufficient
to take
$T^{\mu}$
a tri-linear expression in the fields 
$
H_{\mu\nu}, \Phi, u_{\mu}, \tilde{u}_{\mu}, A_{\mu}
$
verifying the following conditions:
\bea
U_{g} T^{\mu} U^{-1}_{g} = \Lambda^{\mu\nu} T_{\nu}, 
\quad \forall g \in {\cal P}
\nonumber \\
gh(T) = 1
\nonumber \\
3 \leq deg(T^{\mu}) \leq 5.
\eea

If we define
\be
\tilde{T}^{\mu} \equiv T^{\mu} - X^{\mu}
\ee
then we get from (\ref{gauge1}) and (\ref{dQT}):
\be
u^{\mu}~Y_{\mu} + Z_{4} + Z_{6} = \partial_{\mu}\tilde{T}^{\mu}.
\label{gauge}
\ee

The generic form for
$
\tilde{T}^{\mu}
$
is
\be
\tilde{T}^{\mu} = u_{\nu}~T^{\mu\nu} + (\partial_{\rho}u_{\nu})~T^{\mu\nu\rho} 
+ (\partial_{\rho}\partial_{\sigma}u_{\nu})~T^{\mu\nu\rho\sigma}
+ d_{0} u^{\mu} u^{\nu} \tilde{u}_{\nu}
+ d_{1} \epsilon^{\mu\nu\rho\sigma} u_{\nu} u_{\rho} \tilde{u}_{\sigma} 
+ S^{\mu} 
\label{tmu}
\ee
where the expressions
$
T^{\mu\nu}, T^{\mu\nu\rho}, T^{\mu\nu\rho\sigma} 
$
are bi-linear in the fields
$
H_{\mu\nu}, \Phi, A_{\mu},
$
the expression
$
S^{\mu} 
$
is tri-linear in the ghost fields and of canonical dimension $5$
and
$d_{0}, d_{1}$
are constants. We have by direct computation:
\bea
\partial_{\mu}\tilde{T}^{\mu} 
= u_{\nu}~(\partial_{\mu}T^{\mu\nu}) + 
(\partial_{\rho}u_{\nu})~( T^{\rho\nu} + \partial_{\mu}T^{\mu\nu\rho}) 
+ (\partial_{\rho}\partial_{\sigma}u_{\nu})~
(\partial_{\mu}T^{\mu\nu\rho\sigma} + T^{\sigma\nu\rho}) 
\nonumber \\
+ (\partial_{\mu}\partial_{\rho}\partial_{\sigma}u_{\nu})~T^{\mu\nu\rho\sigma} 
+ d_{0} \left[ (\partial_{\mu}u^{\mu}) u^{\nu} \tilde{u}_{\nu} 
+ u^{\mu} (\partial_{\mu}u^{\nu}) \tilde{u}_{\nu} 
+ u^{\mu} u^{\nu} (\partial_{\mu}\tilde{u}_{\nu}) \right] 
\nonumber \\
+ d_{1} \epsilon^{\mu\nu\rho\sigma} 
\left[ 2 (\partial_{\mu}u_{\nu}) u_{\rho} \tilde{u}_{\sigma} 
+ u_{\nu} u_{\rho} (\partial_{\mu}\tilde{u}_{\sigma}) \right] 
+ \partial_{\mu}S^{\mu}.
\label{dtmu}
\eea

Let us write
\be
T^{\mu\nu\rho\sigma} = T^{\mu\nu\rho\sigma}_{1} + T^{\mu\nu\rho\sigma}_{2}
\ee
where the tensor
$
T^{\mu\nu\rho\sigma}_{1}
$
does not contain terms with the factors
$
\eta^{\mu\rho}, \eta^{\mu\sigma}, \eta^{\rho\sigma} 
$
and
$
T^{\mu\nu\rho\sigma}_{2}
$
collects all terms containing at least one of these factors. The terms with
the factor
$\eta^{\mu\rho}$
can be discarded if we redefine
$
T^{\mu\nu}.
$
It follows that the generic form is
\be
T^{\mu\nu\rho\sigma}_{2} = {1 \over 2} ( \eta^{\mu\rho} T^{\nu\sigma}_{2}
+ \eta^{\mu\sigma} T^{\nu\rho}_{2} ).
\ee
We also write
\be
T^{\mu\nu\rho} = T^{\mu\nu\rho}_{1} + \eta^{\mu\rho} T^{\nu}
\ee
where
$
T^{\mu\nu\rho}_{1}
$
collects all terms without the factor
$
\eta^{\mu\rho}.
$

Then we obtain from (\ref{dtmu})
\bea
\partial_{\mu}\tilde{T}^{\mu} 
= u_{\nu}~(\partial_{\mu}T^{\mu\nu} - m^{2} T^{\nu}) + 
(\partial_{\rho}u_{\nu})~( T^{\rho\nu} + \partial_{\mu}T^{\mu\nu\rho}_{1}
+ \partial^{\rho}T^{\nu} - m^{2} T^{\rho\nu}_{2} )
\nonumber \\
+ (\partial_{\rho}\partial_{\sigma}u_{\nu})~
(\partial_{\mu}T^{\mu\nu\rho\sigma}_{1} + T^{\sigma\nu\rho}_{1}
+ \partial^{\rho}T^{\sigma\nu}_{2} )
+ (\partial_{\mu}\partial_{\rho}\partial_{\sigma}u_{\nu})~
T^{\mu\nu\rho\sigma}_{1} 
\nonumber \\
+ d_{0} \left[ (\partial_{\mu}u^{\mu}) u^{\nu} \tilde{u}_{\nu} 
+ u^{\mu} (\partial_{\mu}u_{\nu}) \tilde{u}^{\nu} 
+ u^{\mu} u^{\nu} (\partial_{\mu}\tilde{u}_{\nu}) \right] 
\nonumber \\
+ d_{1} \epsilon^{\mu\nu\rho\sigma} 
\left[ 2 (\partial_{\mu}u_{\nu}) u_{\rho} \tilde{u}_{\sigma} 
+ u_{\nu} u_{\rho} (\partial_{\mu}\tilde{u}_{\sigma}) \right]
+ \partial_{\mu}S^{\mu}.
\eea

It follows that the equation (\ref{gauge}) is equivalent to the following
system:
\bea
{\cal S}_{\mu\rho\sigma}(T^{\mu\nu\rho\sigma}_{1}) = 0
\nonumber \\
T^{\mu\nu\rho}_{1+} = - \partial_{\sigma}T^{\sigma\nu\rho\mu}_{1}  
- {1 \over 2} \left( \partial^{\rho}T^{\mu\nu}_{2} +
\partial^{\mu}T^{\rho\nu}_{2} \right)
\nonumber \\
T^{\rho\nu} = - \partial_{\mu}T^{\mu\nu\rho}_{1} - \partial^{\rho}T^{\nu} 
+ m^{2} T^{\rho\nu}_{2}
\nonumber \\
Y^{\nu} = \partial_{\mu}T^{\mu\nu} - m^{2} T^{\nu}
\nonumber \\
d_{Q} \left[ (\partial_{\mu}u^{\mu}) u^{\nu} \tilde{u}_{\nu} 
+ u^{\mu}) (\partial_{\mu}u_{\nu}) \tilde{u}^{\nu} 
+ u^{\mu} u^{\nu} (\partial_{\mu}\tilde{u}_{\nu}) \right] 
\nonumber \\
+ d_{1} \epsilon^{\mu\nu\rho\sigma} 
\left[ 2 (\partial_{\mu}u_{\nu}) u_{\rho} \tilde{u}_{\sigma} 
+ u_{\nu} u_{\rho} (\partial_{\mu}\tilde{u}_{\sigma}) \right]
+ (\partial_{\mu}S^{\mu})_{4} = Z_{4}
\nonumber \\
(\partial_{\mu}S^{\mu})_{6} = Z_{6}.
\label{gaugeA}
\eea

Here we have written
\be
T^{\mu\nu\rho}_{1} = T^{\mu\nu\rho}_{1+} + T^{\mu\nu\rho}_{1-}
\ee
where the two pieces have the following symmetry properties:
\be
T^{\mu\nu\rho}_{1\epsilon} = \epsilon T^{\rho\nu\mu}_{1\epsilon} \quad
\forall \epsilon = \pm.
\ee

The expression
$
(\partial_{\mu}S^{\mu})_{j}
$
contains the terms of canonical dimension $j = 4, 6$ of
$
\partial_{\mu}S^{\mu};
$
the expression
$
(\partial_{\mu}S^{\mu})_{4}
$
is proportional to the mass, so is zero in the massless case.

From the first four equations of the system (\ref{gaugeA}) we obtain
\be
Y^{\nu} = (\partial^{2} + m^{2}) (\partial_{\mu}T^{\mu\nu}_{1} - T^{\nu}),
\ee
so
$
Y_{\mu}
$
has the generic form
\be
Y_{\mu} = {1\over 2} (\partial^{2} + m^{2}) Z_{\mu}
\label{yz}
\ee
where
$
Z_{\mu}
$
is a Wick polynomial bilinear in the fields
$
H_{\mu\nu}, \Phi, A_{\mu}
$
and verifies the following conditions:
\bea
U_{g} Z^{\mu} U^{-1}_{g} = \Lambda^{\mu\nu} Z_{\nu}, 
\quad \forall g \in {\cal P}
\nonumber \\
gh(T) = 0
\nonumber \\
2 \leq deg(T^{\mu}) \leq 3.
\eea

The generic form of
$
Z_{\mu}
$
is
\bea
Z_{\mu} = f_{1} \Phi (\partial_{\mu}\Phi) 
+ f_{2} H^{\alpha\beta} (\partial_{\mu}H_{\alpha\beta})
+ f_{3} \Phi (\partial^{\nu}H_{\mu\nu})
+ f_{4} H_{\mu\nu} (\partial^{\nu}\Phi)
+ f_{5} H_{\mu\nu} (\partial_{\rho}H^{\nu\rho})
\nonumber \\
+ f_{6} H^{\nu\rho} (\partial_{\rho}H_{\mu\nu})
+ f_{7} \epsilon_{\mu\nu\alpha\beta} H^{\alpha\rho}
(\partial^{\nu}H^{\beta}_{\cdot\rho})
+ f_{8} \Phi v_{\mu} + f_{9} H_{\mu\nu} v^{\nu}
+ f_{10} v_{\mu} (\partial_{\alpha}v_{\alpha})
\nonumber \\
+ f_{11} v^{\alpha} (\partial_{\mu}v_{\alpha})
+ f_{12} v^{\alpha} (\partial_{\alpha}v^{\mu})
+ f_{13} \epsilon_{\mu\nu\alpha\beta} v^{\alpha} (\partial^{\nu}v^{\beta}).
\eea

The basic equation (\ref{yz}) becomes now
\bea
Y_{\mu} = f_{1} (\partial^{\nu}\Phi) (\partial_{\mu}\partial_{\nu}\Phi) 
+ f_{2} (\partial^{\nu}H^{\alpha\beta}) 
(\partial_{\mu}\partial_{\nu}H_{\alpha\beta})
+ f_{3} (\partial_{\rho}\Phi) (\partial^{\nu}\partial^{\rho}H_{\mu\nu})
+ f_{4} (\partial_{\rho}H_{\mu\nu}) (\partial^{\nu}\partial^{\rho}\Phi)
\nonumber \\
+ f_{5} (\partial^{\sigma}H_{\mu\nu}) 
(\partial_{\rho}\partial_{\sigma}H^{\nu\rho})
+ f_{6} (\partial^{\sigma}H^{\nu\rho}) 
(\partial_{\rho}\partial_{\sigma}H_{\mu\nu})
+ f_{7} \epsilon_{\mu\nu\alpha\beta} (\partial^{\sigma}H^{\alpha\rho})
(\partial^{\nu}\partial_{\sigma}H^{\beta}_{\cdot\rho})
\nonumber \\
+ f_{8} (\partial^{\nu}\Phi) (\partial_{\nu}v_{\mu}) 
+ f_{9} (\partial_{\rho}H_{\mu\nu}) (\partial^{\rho}v^{\nu})
+ f_{10} (\partial_{\nu}v_{\mu}) (\partial^{\nu}\partial_{\alpha}v_{\alpha})
\nonumber \\
+ f_{11} (\partial^{\nu}v^{\alpha}) (\partial_{\mu}\partial_{\nu}v_{\alpha})
+ f_{12} (\partial^{\nu}v^{\alpha}) (\partial_{\nu}\partial_{\alpha}v^{\mu})
+ f_{13} \epsilon_{\mu\nu\alpha\beta} (\partial^{\rho}v^{\alpha}) 
(\partial_{\rho}\partial^{\nu}v^{\beta})
 - {1 \over 2} m^{2} Z_{\mu}.
\label{yz1}
\eea

If we substitute here the expressions (\ref{y}) - (\ref{y22}) we get the
a system of equations of $78$ equations for $73$ the unknowns
$c^{(j)}, d^{(j)}, f_{j}$
($37 + 23 + 13 = 73$).
One can solve this system explicitly, the only non-zero coefficients are:
\bea
c^{(1)} = {2\over 3} m^{2} a  - {1\over 3} m^{2} b \quad
c^{(2)} = {1\over 2} m^{2} a \quad 
c^{(3)} = - {3\over 4} m^{2} a \quad 
c^{(4)} = - m^{2} b 
\quad
c^{(5)} = - m^{2} a. 
\nonumber \\
c^{(6)}_{1} = - 2 a \quad c^{(6)}_{6} = - 4 a \quad 
c^{(7)}_{2} = 2 a \quad c^{(8)} = 4 a  \quad c^{(9)} = a 
\nonumber \\
\quad c^{(10)} = a \quad c^{(11)}_{2} = 4 a \quad c^{(11)}_{3} = 4 a  
\quad c^{(13)}_{3} = 4 a \quad c^{(14)}_{2} = 2 a  
\nonumber \\
d^{(1)} = {3\over 2} m^{2} b \quad
d^{(2)} = m^{2} a \quad    
d^{(3)}_{1} = m b \quad
d^{(3)}_{3} = - 2 m b \quad
d^{(4)} = m b 
\nonumber \\    
d^{(7)}_{1} = m b \quad
d^{(7)}_{2} = - 4 m a - {1\over 2} m b \quad
d^{(7)}_{3} = m b \quad
d^{(8)}_{1} = m b \quad
d^{(8)}_{2} = m b \quad
d^{(8)}_{3} = - { 1\over 2} m b \quad
\nonumber \\
d^{(9)} = 2 m b \quad
d^{(10)}_{1} = 4 a - b \quad
d^{(10)}_{2} = 2 b \quad
d^{(10)}_{3} = - 2 b 
\eea
and
\bea
f_{1} = - {1 \over 2} a \quad f_{2} = - a \quad 
f_{4} = - 2 a \quad f_{5} = 4 a  \quad f_{6} = 4 a; 
\nonumber \\
f_{9} = - m b \quad
f_{10} = b \quad
f_{11} = 2 a - {1\over 2} b \quad
f_{12} = - b.
\eea
here 
$a, b \in \R$
are arbitrary parameters. Now we can determine the expression
$Z_{6}$ 
i.e. the terms tri-linear in the ghost fields and of canonical dimension $5$:
\be
Z_{6} = 4a~u^{\alpha} (\partial^{\mu}u^{\nu}) 
(\partial_{\mu}\partial_{\nu}\tilde{u}_{\alpha})
\label{z6}
\ee
and the last equation of the system (\ref{gaugeA}) admits a (non-unique)
solution. One can choose the ``minimal" solution:
\be
S^{\mu} = 2a \left[ 
u_{\nu} (\partial^{\mu}u_{\rho}) (\partial^{\nu}\tilde{u}^{\rho})
+ u_{\nu} (\partial^{\mu}u_{\rho}) (\partial^{\rho}\tilde{u}^{\nu})
- u_{\nu} u_{\rho} (\partial^{\mu}\partial^{\nu}\tilde{u}^{\rho}) \right].
\label{smu}
\ee

Finally we consider the fifth equation (\ref{gaugeA}). First we have
the explicit expression:
\bea
Z_{4} \equiv  
- {1 \over 2} \left[ c^{(4)} + m d^{(7)}_{2} \right] 
(\partial^{\mu}u^{\nu}) u_{\mu} \tilde{u}_{\nu}
- {1 \over 2} \left[ c^{(4)} + m d^{(7)}_{3} \right] 
(\partial^{\nu}u^{\mu}) u_{\mu} \tilde{u}_{\nu}
\nonumber \\
+ {1 \over 2} \left[ {1 \over 2} c^{(4)} + c^{(5)} 
- {1 \over 4} m^{2} c^{(11)}_{2} - {1 \over 2} m^{2} c^{(13)}_{3} 
- m d^{(7)}_{1} \right] 
(\partial^{\nu}u_{\nu}) u_{\mu} \tilde{u}^{\mu}
\nonumber \\
+ {1 \over 2} \left[ m^{2} c^{(11)}_{2} - m^{2} c^{(11)}_{3} 
- m d^{(8)}_{1} + m d^{(8)}_{3} \right] 
u^{\mu} u^{\nu} (\partial_{\mu}\tilde{u}_{\nu}).
\label{z4}
\eea

It follows easily that we can take
$
d_{1} = 0
$
in the expression (\ref{tmu}) of
$\tilde{T}^{\mu}$.
Then we get from the fifth equation (\ref{gaugeA}):
\be
d_{0} = - 2 m^{2} a - {3\over 4} m^{2} b,
\ee
so we get the result from the statement. 
$\qed$

We remark that the basic idea of solving the gauge invariance problem was to 
use the equation (\ref{yz}) instead of the original gauge invariance condition 
(\ref{gauge1}); the equation (\ref{yz}) is simpler because we do not need an
ansatz for
$T^{\mu}$; 
only an ansatz for
$Z^{\mu}$
is necessary.

The solution 
$T^{(a)}$ 
gives in the massless limit the usual gravity theory \cite{Sc2} plus the new
term
$
4 H^{\mu\nu} (\partial_{\mu}v_{\rho}) (\partial_{\nu}v^{\rho})
.
$
The usual choice is
$a = - {1\over 4}$.

One can re-express 
$T^{(a)}$ 
using the
 variables
 from (\ref{newH}); we have
\begin{prop}
In the variables
\be
h_{\mu\nu} \equiv H_{\mu\nu} + {1\over 2} \eta_{\mu\nu} \Phi
 \qquad
h \equiv h^{\mu}_{\cdot\mu}
 = 2 \Phi
\ee
the expression 
$T^{(a)}$ 
from the preceding theorem is equivalent to:
\bea
T = h^{\mu\nu} (\partial_{\mu}h) (\partial_{\nu}h)
- 2 h^{\mu\nu} (\partial_{\mu}h_{\rho\sigma}) (\partial_{\nu}h^{\rho\sigma})
- 4 h_{\mu\nu} (\partial_{\rho}h^{\mu\sigma}) (\partial_{\sigma}h^{\nu\rho})
\nonumber \\
- 2 h^{\mu\nu} (\partial_{\rho}h_{\mu\nu}) (\partial^{\rho}h)
+ 4 h^{\mu\nu} (\partial_{\sigma}h_{\mu\rho}) 
(\partial^{\sigma}h_{\nu}^{\cdot\rho})
\nonumber \\
+ 4 (\partial_{\mu}h^{\mu\nu}) u^{\rho} (\partial_{\rho}\tilde{u}_{\nu})
- 4 h^{\mu\nu} (\partial_{\mu}u^{\rho}) (\partial_{\nu}\tilde{u}_{\rho})
+ 4 h^{\mu\nu} (\partial_{\mu}v_{\rho}) (\partial_{\nu}v^{\rho})
\nonumber \\
- 4 m (\partial_{\mu}v_{\nu}) u^{\mu} \tilde{u}^{\nu}
+ m^{2} \left( - {4 \over 3} h^{\mu\nu} h_{\mu\rho} h_{\nu}^{\cdot\rho}
+ h^{\mu\nu} h_{\mu\nu} h
 -{1 \over 6} h^{3}
 \right).
\label{first-order} 
\eea

In these conditions one can take in (\ref{gauge1})
\bea
T^{(a)}_{\mu} = u^{\nu} [ - (\partial_{\mu}h) (\partial_{\nu}h)
+ 2 (\partial_{\mu}h_{\rho\sigma}) (\partial_{\nu}h^{\rho\sigma})
-  4 (\partial_{\rho}h^{\rho\sigma}) (\partial_{\nu}h_{\mu\sigma})
\nonumber \\
+ 4 (\partial_{\nu}h^{\rho\sigma}) (\partial_{\rho}h_{\mu\sigma})
- 4 (\partial_{\mu}v_{\rho}) (\partial_{\nu}v^{\rho})
 ]
\nonumber \\
+ u_{\mu} \left[ {1\over 2} (\partial_{\nu}h) (\partial^{\nu}h)
- (\partial_{\nu}h_{\rho\sigma}) (\partial^{\nu}h^{\rho\sigma})
- 2 (\partial_{\rho}h_{\nu\sigma}) (\partial^{\sigma}h^{\nu\rho})
+ 2 (\partial_{\nu}v_{\rho}) (\partial^{\nu}v^{\rho})
 \right]
\nonumber \\
+ (\partial^{\rho}u^{\nu}) [ 4 h_{\rho\sigma} (\partial^{\sigma}h_{\mu\nu})
+ 4 h_{\mu\sigma} (\partial_{\nu}h^{\sigma}_{\cdot\rho})
+ 2 h_{\nu\rho} (\partial_{\mu}h) 
- 4 h_{\rho\sigma} (\partial_{\mu}h^{\sigma}_{\cdot\nu}) ]
\nonumber \\
- 4 (\partial^{\rho}u_{\mu}) h^{\nu\sigma} (\partial_{\nu}h_{\rho\sigma})
+ (\partial^{\rho}u_{\rho}) [ - h (\partial_{\mu}h)
+ 2 h^{\nu\sigma} (\partial_{\mu}h_{\nu\sigma}) ] 
\nonumber \\
+ 2 [ u^{\nu} (\partial_{\mu}u^{\rho}) (\partial_{\nu}\tilde{u}_{\rho})
+ u^{\nu} (\partial_{\nu}u^{\rho}) (\partial_{\mu}\tilde{u}_{\rho})
- u_{\mu} (\partial^{\nu}u^{\rho}) (\partial_{\nu}\tilde{u}_{\rho}) ]
\nonumber \\
+ m^{2} u_{\mu} \left( h^{\nu\rho} h_{\nu\rho} - {1\over 2} h^{2} \right).
\eea
\label{Ta}
\end{prop}

{\bf Proof:}
It is convenient to start from the expression $T$ above and make the 
substitution
$
H_{\mu\nu} \equiv h_{\mu\nu} - {1\over 2} \eta_{\mu\nu} \Phi
 \qquad
h \equiv 2 \Phi.
$
Then one makes the transformations described at the beginning of the proof of
the preceding theorem, namely we get rid of derivatives appearing on
$u_{\rho}$
subtraction total divergences. We also note that the fourth and the fifth terms
from the expression of $T$ can be eliminated if we use the identity 
(\ref{magic}) but have been included such that the linear approximation of the
Hilbert Lagrangian is reproduced \cite{Sc2}. We obtain the expression
$T^{(a)}$
from the statement of the theorem.

The computation of
$T^{(a)}_{\mu}$
is not very difficult and provides an impressive check that the preceding
computations are right. It is not hard to obtain by direct computation that
\be
d_{Q}T = i [ u_{\mu} t^{\mu} + (\partial_{\nu}u_{\mu}) t^{\mu\nu}
+ (\partial_{\nu}\partial_{\rho}u_{\mu}) t^{\mu\nu\rho} + s ]
\ee
where the expressions
$
t^{\mu}, t^{\mu\nu}, t^{\mu\nu\rho}
$
are bi-linear in the fields
$h_{\mu\nu}, v_{\mu}$
and $s$ is tri-linear in the ghost fields. 
We note that we have a certain freedom in choosing the expression
$
t^{\mu\nu\rho}
$
if we do not impose symmetry in 
$\nu, \rho$.
Now, as in the proof of the preceding theorem, one makes ``integrations by 
parts" and rewrites the preceding expressions as follows: 
\be
d_{Q}T = i ( u^{\mu} y_{\mu} + \partial^{\mu}x_{\mu} + s )
\ee
where
\be
y_{\mu} = t_{\mu} - \partial^{\nu}t_{\mu\nu} 
+ \partial^{\nu}\partial^{\rho}t_{\mu\nu\rho};
\ee
and
\bea
x_{\mu} = u^{\nu} x_{\mu\nu} + (\partial^{\rho}u^{\nu}) x_{\mu\nu\rho} 
\nonumber \\
x_{\mu\nu} \equiv t_{\nu\mu} - \partial^{\rho}t_{\nu\rho\mu} \qquad
x_{\mu\nu\rho} \equiv t_{\nu\mu\rho}.
\eea

By direct computation we can prove that
\be
y_{\mu} = 0
\ee
It remains to prove that the expression 
$s_{\mu}$
from the statement is such that
$s = \partial^{\mu}s_{\mu}$
and the expression
$d_{Q}T$
above is exhibited as the total divergence from the statement. 
$\qed$

In the same way one can also re-express 
$T^{(b)}$ 
using the
 variables
 from (3.60); we have
\begin{prop}
The expression 
$T^{(b)}$ 
from the preceding theorem is equivalent to:
\bea
T = - h^{\mu\nu} (\partial_{\mu}v_{\rho}) (\partial_{\nu}v^{\rho})
+ 2 h^{\mu\nu} (\partial_{\mu}v_{\nu}) (\partial^{\rho}v_{\rho})
- {1\over 2} v (\partial_{\rho}v^{\rho})^{2}
\nonumber \\
- 2 h^{\mu\nu} (\partial_{\mu}v_{\rho}) (\partial^{\rho}v_{\nu})
+ {1\over 2} h (\partial_{\mu}v_{\rho})
 (\partial^{\rho}v^{\mu})
\nonumber \\
+ m \Bigl[ h^{\mu\nu} (\partial_{\rho}h_{\mu\nu}) v^{\rho}
- {1\over 2} h (\partial_{\rho}h) v^{\rho} 
- 2 h^{\mu\nu} (\partial_{\nu}h_{\mu\rho}) v^{\rho}
+ h_{\rho\nu} (\partial^{\nu}h) v^{\rho}
\nonumber \\
+ {1\over 2} h (\partial^{\mu}h_{\mu\rho}) v^{\rho}
+ (\partial_{\rho}v^{\rho}) u_{\mu} \tilde{u}^{\mu}
- {1 \over 2} (\partial_{\mu}v_{\nu}) u^{\mu} \tilde{u}^{\nu}
+ (\partial_{\nu}v_{\mu}) u^{\mu} \tilde{u}^{\nu}
+ v^{\nu} u^{\mu} (\partial_{\nu}\tilde{u}_{\mu})
\nonumber \\
+ v_{\mu} u^{\mu} (\partial^{\nu}\tilde{u}_{\nu})
- {1 \over 2} v^{\nu} u^{\mu} (\partial_{\mu}\tilde{u}_{\nu})
+ 2 v_{\mu} v_{\nu} (\partial^{\mu}v^{\nu}) \Bigl]
\nonumber \\
+ m^{2} \Bigl( - {1 \over 3} h^{\mu\nu} h_{\mu\rho} h_{\nu}^{\cdot\rho}
+ {1\over 4} h^{\mu\nu} h_{\mu\nu} h
 - {1 \over 24} h^{3}
- h^{\mu\nu} u_{\mu} \tilde
{u}_{\nu}
\nonumber \\
+ {1\over 4} h u^{\mu} \tilde
{u}_{\mu}
+ {3 \over 2} h^{\mu\nu} v_{\mu} v_{\nu}
- {1 \over 4} h v^{\mu} v_{\mu}
 \Bigl)
\eea

In these conditions one can take in (\ref{gauge1})
\bea
T^{(b)}_{\mu} = u^{\nu} \Bigl\{ 
(\partial_{\mu}v_{\rho}) (\partial_{\nu}v^{\rho})
- (\partial_{\nu}v_{\mu}) (\partial_{\rho}v^{\rho})
- {1\over 2} (\partial_{\mu}v_{\nu}) (\partial_{\rho}v^{\rho})
\nonumber \\
+ (\partial_{\nu}v_{\rho}) (\partial^{\rho}v_{\mu})
+ {1\over 2} (\partial_{\mu}v_{\rho}) (\partial^{\rho}v_{\nu})
+ {1\over 2} v_{\nu} (\partial_{\mu}\partial_{\rho}v^{\rho})
- {1\over 2} v^{\rho} (\partial_{\mu}\partial_{\rho}v_{\nu})
\nonumber \\
+ m \Bigl[ - (\partial^{\rho}h_{\mu\rho}) v_{\nu}
- h_{\mu\nu} (\partial^{\rho}v_{\rho})
+ h_{\mu\rho} (\partial_{\nu}v^{\rho})
+ h_{\nu\rho} (\partial^{\rho}v_{\mu})
- {1\over 2} v (\partial_{\nu}v_{\mu})
\nonumber \\
- (\partial_{\rho}h_{\mu\nu}) v^{\rho}
+ (\partial_{\nu}h_{\mu\rho}) v^{\rho}
+ {1\over 2} (\partial_{\mu}h_{\nu\rho}) v^{\rho}
- {1\over 4} (\partial_{\mu}h) v_{\nu}
\nonumber \\
- {1\over 2} (\partial_{\nu}h) v_{\mu}
+ {1\over 4} h (\partial_{\mu}v_{\nu}) 
- {1\over 2} h_{\nu\rho} (\partial_{\mu}v^{\rho})
\Bigl]
\nonumber \\
+ m^{2} \Bigl( - {5\over 2} v_{\mu} v_{\nu}
+ h_{\mu\rho} h_{\nu}^{\cdot\rho}
- {1\over 2} h_{\mu\nu} h \Bigl) \Bigl\}
\nonumber \\
+ u_{\mu} \Bigl\{ 
- {1\over 2} (\partial_{\nu}v_{\rho}) (\partial^{\nu}v^{\rho})
+ {1\over 2} (\partial_{\rho}v_{\rho})^{2}
- {1\over 2} (\partial_{\nu}v_{\rho}) (\partial^{\rho}v^{\nu})
\nonumber \\
+ m \Bigl[ - h^{\rho\sigma} (\partial_{\rho}v_{\sigma})
+ {1\over 2} h (\partial_{\rho}v^{\rho})
+ {1\over 2} (\partial_{\rho}h) v^{\rho}
- {1\over 2} (\partial^{\rho}h_{\rho\sigma}) v^{\sigma} \Bigl]
\nonumber \\
+ m^{2} \Bigl( - {1 \over 4} h_{\rho\sigma} h^{\rho\sigma}
+ {1\over 8} h^{2}
+ {1 \over 2} v^{\nu} v_{\nu}
\Bigl ) \Bigl\}
+ m (\partial^{\rho}u^{\nu}) h_{\mu\rho} v_{\nu}
\nonumber \\
- {1\over 2} (\partial_{\mu}u_{\nu}) \Bigl[ v^{\nu} (\partial^{\rho}v_{\rho})
- v_{\rho} (\partial^{\rho}v^{\nu})
- m \Bigl( h^{\nu\rho} v_{\rho} - {1\over 2} h v^{\nu} \Bigl) \Bigl] 
- {3\over 4} m^{2} u_{\mu} u_{\nu} \tilde{u}^{\nu}.
\label{Tb}
\eea
\end{prop}

\newpage
\section{Second Order Gauge Invariance\label{second}}

In this Section we consider the second order gauge invariance. For this we
must construct the chronological products
$T(x,y)$
and
$T_{\mu}(x,y)$
such that the identity (\ref{gauge}) is verified. The construction
procedure is well-known: one computes first the corresponding causal 
commutators
$[ T(x), T(y)]$
and
$[ T_{\mu}(x), T(y) ]$
and makes the substitution
$D_{m}(x-y) \mapsto D^{F}_{m}(x-y)$
i.e. one substitutes the causal Pauli-Jordan distribution by the corresponding
Feynman propagator and similar substitutions for the loop graphs; one obtains 
the expressions
$T^{F}(x,y)$
and
$T_{\mu}^{F}(x,y)$
which verify all Bogoliubov axioms but might spoil second order gauge 
invariance. To restore it we must annihilate some anomalies and make finite
renormalizations. These finite renormalizations must also preserve the
power counting theorem which in this case says \cite{Sc2} that the expressions
$T(x,y)$
and
$T^{\mu}(x,y)$
should be of the form
\be
T(x,y) = \sum_{j} t_{j}(x-y) W_{j}(x,y)
\label{generic}
\ee
where
$W_{j}$
are Wick polynomials and
$t_{j}$
are distributions such that one has
\be
\omega(t_{j}) + deg(W_{j}) \leq 6.
\label{power}
\ee

The origin of the anomalies is explained in \cite{Gr2}, \cite{Sc2}. One
starts from the identity
\be
d_{Q} [ T(x),T(y) ] = 
i {\partial \over \partial x^{\mu}} [ 
T^{\mu}(x),T(y)]
+ i {\partial \over \partial y^{\mu}} [ T(x),
T^{\mu}(y)]
\ee
which follows from first order gauge invariance. If one substitutes expressions
of the type (\ref{generic}) for the causal commutators
\be
D(x,y) = \sum_{j} d_{j}(x-y) W_{j}(x,y)
\ee
then the preceding identity reduces to some identities verified by the 
distributions
$
d_{j}.
$
When we make the causal splitting  of the these distributions, preserving the 
degree of singularity, some of these identities are lost and we get anomalies. 
From tree Feynman graphs we get the identity
\be
(\partial^{2} + m^{2} ) D_{m} = 0,
\label{G1}
\ee
which cannot be split causally preserving the degree of singularity; indeed
it is well known that
\be
(\partial^{2} + m^{2} ) D^{F}_{m} = \delta (x-y).
\label{W1}
\ee

From loop graphs we get the identities
\be
\partial_{\mu}D_{m_{1},m_{2}}^{\mu} = - m_{2}^{2} D_{m_{1},m_{2}}
+ \eta_{\mu\nu} D_{m_{1},m_{2}}^{\mu\nu}
\label{G2}
\ee
\be
\partial_{\mu}D_{m_{1},m_{2}}^{\mu\nu} = - m_{1}^{2} D_{m_{2},m_{1}}^{\nu}
+ \tilde{D}_{m_{1},m_{2}}^{\mu}
\label{G3}
\ee
\be
\partial_{\mu}\tilde{D}_{m_{1},m_{2}}^{\mu\nu} 
= - m_{2}^{2} D_{m_{1},m_{2}}^{\nu}
+ \tilde{D}_{m_{1},m_{2}}^{\mu}
\label{G4}
\ee
where we have introduced the following distributions with causal support:
\bea
D_{m_{1},m_{2}} \equiv D^{(+)}_{m_{1}} D^{(+)}_{m_{2}} -
D^{(-)}_{m_{1}} D^{(-)}_{m_{2}}
\nonumber \\
D_{m_{1},m_{2}}^{\mu} \equiv D^{(+)}_{m_{1}} \partial^{\mu}D^{(+)}_{m_{2}} 
- D^{(-)}_{m_{1}} \partial^{\mu}D^{(-)}_{m_{2}} 
\nonumber \\
D_{m_{1},m_{2}}^{\mu\nu} \equiv 
\partial^{\mu}D^{(+)}_{m_{1}} \partial^{\nu}D^{(+)}_{m_{2}}
- \partial^{\mu}D^{(-)}_{m_{1}} \partial^{\nu}D^{(-)}_{m_{2}}
\nonumber \\
\tilde{D}_{m_{1},m_{2}}^{\mu\nu} \equiv 
D^{(+)}_{m_{1}} \partial^{\mu}\partial^{\nu}D^{(+)}_{m_{2}}
- D^{(-)}_{m_{1}} \partial^{\mu}\partial^{\nu}D^{(-)}_{m_{2}}.
\nonumber \\
\tilde{D}_{m_{1},m_{2}}^{\mu} \equiv 
\partial_{\nu}D^{(+)}_{m_{1}} \partial^{\mu}\partial^{\nu}D^{(+)}_{m_{2}}
- \partial_{\nu}D^{(-)}_{m_{1}} \partial^{\mu}\partial^{\nu}D^{(-)}_{m_{2}}.
\eea
It is easy to prove that one can use the arbitrariness of the causal
splitting of these distributions such that one can eliminate all anomalies
of the Ward identities (\ref{G2}) - (\ref{G4}). So it follows that only
tree Feynman graphs can produce anomalies. 

We apply this strategy to the quantum gravity model from the preceding Section.

First we consider the theory given by the interaction Lagrangian
$T = T^{(a)}$.
\begin{thm}
The Lagrangian
$T = T^{(a)}$
gives a theory gauge invariant in the second order of the perturbation theory
if we perform convenient finite renormalizations of the second-order
chronological products
\be
T(x,y) = T^{F}(x,y) + i~\delta(x - y) N(x) 
\qquad
T_{\mu}(x,y) = T^{F}_{\mu}(x,y) + i~\delta(x - y) i~N_{\mu}(x) 
\ee
where $N$ and 
$N^{\mu}$
are some Wick polynomials. In particular:
\bea
N = 16 h^{\mu\nu} h^{\rho\sigma} 
(\partial_{\rho}h_{\mu\nu}) (\partial_{\sigma}h)
- 8 h^{\mu\nu} h^{\rho\sigma} 
(\partial_{\alpha}h_{\mu\nu}) (\partial^{\alpha}h_{\rho\sigma})
- 32 h^{\mu\nu} h_{\nu\rho} 
(\partial^{\alpha}h^{\rho\beta}) (\partial_{\beta}h_{\mu\alpha})
\nonumber \\
- 32 h^{\mu\nu} h^{\rho\sigma} 
(\partial_{\mu}h^{\rho\alpha}) ({\partial_{\nu}h_{\sigma}}^{\cdot\alpha})
+ 32 h^{\mu\nu} h_{\nu\rho} 
(\partial^{\alpha}h_{\mu\beta}) (\partial_{\alpha}h^{\rho\beta})
+ 16 h^{\mu\nu} h^{\rho\sigma} 
(\partial_{\alpha}h_{\mu\rho}) (\partial^{\alpha}h_{\nu\sigma})
\nonumber \\
- 16 h^{\mu\nu} h_{\nu\rho} 
({\partial^{\alpha}h_{\mu}}^{\cdot\rho}) (\partial_{\alpha}h)
+ 16 u^{\rho} (\partial_{\rho}\tilde{u}^{\nu}) 
u^{\sigma} (\partial_{\sigma}\tilde{u}_{\nu})
\nonumber \\
+ 2 m^{2} \Bigl( {1\over 12} h^{4} - h^{\mu\nu} h_{\mu\nu} h^{2}
+ {8 \over 3} h^{\mu\nu} h_{\nu\rho} h^{\rho}_{\cdot\mu} h
+ h^{\mu\nu} h_{\mu\nu} h^{\rho\sigma} h_{\rho\sigma}
- 4 h^{\mu\nu} h_{\nu\rho} h_{\mu\sigma} h^{\rho\sigma} \Bigl).
\label{second-order}
\eea
\end{thm}

{\bf Proof:} 
As it is known, the first step is to compute the causal commutator
$
[ T^{\mu}(x), T(y) ].
$
The anomalies are produced by two types of terms in
$
T^{\mu}(x)
:
$
(a) with the index $\mu$ appearing in a derivative 
$
\partial_{\mu};
$
(b) with the index $\mu$ appearing in the combination
$h_{\mu\rho}$.
Inspecting the expression from prop. \ref{Ta} we have
\bea
T_{\mu} = T_{1} (\partial_{\mu}h) 
+ T_{2}^{\alpha\beta} (\partial_{\mu}h_{\alpha\beta})
+ T_{3}^{\nu} (\partial_{\mu}\tilde{u}_{\nu}) 
+ T_{4}^{\nu} (\partial_{\mu}u_{\nu}) 
+ T_{5}^{\nu} (\partial_{\mu}v_{\nu})  
\nonumber \\
+ S^{\nu\rho} (\partial_{\nu}h_{\mu\rho}) 
+ S^{\rho} h_{\mu\rho} + \cdots
\eea
where by $\cdots$ we mean terms which do not produce anomalies and we have 
defined:
\bea
T_{1} \equiv - u^{\nu} (\partial_{\nu}h)
 - (\partial^{\rho}u_{\rho}) h
+ 2 (\partial^{\rho}u^{\nu}) h_{\nu\rho}
\nonumber \\
T_{2}^{\alpha\beta} \equiv 2 \Bigl[
u_{\lambda} (\partial^{\lambda}h^{\alpha\beta}) 
+ (\partial^{\lambda}u_{\lambda}) h^{\alpha\beta} 
- (\partial_{\lambda}u^{\alpha}) h^{\lambda\beta} 
- (\partial_{\lambda}u^{\beta}) h^{\lambda\alpha} \Bigl] 
\nonumber \\
T_{3}^{\nu} \equiv 2 u_{\rho} (\partial^{\rho}u^{\nu})
\nonumber \\
T_{4}^{\nu} \equiv - 2 u_{\rho} (\partial^{\rho}\tilde{u}^{\nu})
\nonumber \\
T_{5}^{\nu} \equiv - 4 u_{\rho} (\partial^{\rho}v^{\nu})
\nonumber \\
S^{\nu\rho} = 4 \Bigl[ - u^{\nu} (\partial_{\sigma}h^{\rho\sigma})
+ u_{\sigma} (\partial^{\sigma}h^{\nu\rho})
+ (\partial_{\sigma}u^{\rho}) h^{\nu\sigma} \Bigl] 
\nonumber \\
S^{\rho} = 4 (\partial_{\sigma}u_{\lambda}) 
(\partial^{\lambda}h^{\rho\sigma});
\eea
here we have imposed the symmetry condition
\be
T_{2}^{\rho\sigma} = (\rho \leftrightarrow \sigma).
\ee
It is also convenient to denote
\be
T_{2} \equiv T_{2;\alpha\beta} \eta^{\alpha\beta}
\ee
with the explicit expression
\be
T_{2} = - 2 T_{1}.
\ee 

One has to compute the commutator of
$
\partial^{\mu}h, \partial^{\mu}h^{\alpha\beta}, \partial^{\mu}\tilde{u}^{\rho},
\partial^{\mu}u^{\rho}, \partial^{\mu}A^{\rho}
$
with the 12 linear independent Wick monomials which appear in the expression
of the total coupling $T$. Using the causal (anti)commutation 
relations from Section 4 we get
:
\bea
[ \partial_{\mu}h(x), T(y) ] = i~\partial_{\mu}D_{m}(x-y) D_{1}(y)
+ i~\partial_{\mu}\partial_{\nu}D_{m}(x-y) D^{\nu}_{1}(y)
\nonumber \\
~[ \partial_{\mu}H^{\alpha\beta}(x), T(y) ] 
= i~\partial_{\mu}D_{m}(x-y) D_{2}^{\alpha\beta}(y)
+ i~\partial_{\mu}\partial_{\nu}D_{m}(x-y) D_{2}^{\alpha\beta;\nu}(y)
\nonumber \\
~[ \partial_{\mu}\tilde{u}^{\rho}(x), T(y) ] 
= i~\partial_{\mu}D_{m}(x-y) D_{3}^{\rho}(y)
+ i~\partial_{\mu}\partial_{\nu}D_{m}(x-y) D_{3}^{\rho;\nu}(y)
\nonumber \\
~[ \partial_{\mu}u^{\rho}(x), T(y) ] 
= i~\partial_{\mu}D_{m}(x-y) D_{4}^{\rho}(y)
+ i~\partial_{\mu}\partial_{\nu}D_{m}(x-y) D_{4}^{\rho;\nu}(y)
\nonumber \\
~[ \partial_{\mu}v^{\rho}(x), T(y) ] 
= i~\partial_{\mu}D_{m}(x-y) D_{5}^{\rho}(y)
+ i~\partial_{\mu}\partial_{\nu}D_{m}(x-y) D_{5}^{\rho;\nu}(y)
\eea
where
\bea
D_{1} = - (\partial^{\rho}h) (\partial_{\rho}h)
+ 2 (\partial^{\rho}h^{\alpha\beta}) (\partial_{\rho}h_{\alpha\beta})
- 4 (\partial^{\alpha}h^{\beta\lambda}) (\partial_{\beta}h_{\alpha\lambda})
\nonumber \\ 
- 4 (\partial^{\rho}u^{\sigma}) (\partial_{\rho}\tilde{u}_{\sigma})
+ 4 a~(\partial^{\rho}v^{\sigma}) (\partial_{\rho}v_{\sigma})
\eea
\be
D_{1}^{\nu} = - 4 h^{\nu\rho} (\partial_{\rho}h)
+ 8 h_{\rho\sigma} (\partial^{\rho}h^{\nu\sigma})
+ 2 h (\partial^{\nu}h_{\sigma})
- 4 u_{\rho} (\partial^{\rho}\tilde{u}^{\nu}).
\ee 

\be
D_{2}^{\alpha\beta} = {\cal D}_{2}^{\alpha\beta} + D_{2} \eta^{\alpha\beta},
\ee 
where
\bea
{\cal D}_{2}^{\alpha\beta} = - (\partial^{\alpha}h) (\partial^{\beta}h)
+ 2 (\partial^{\alpha}h^{\sigma\lambda}) (\partial^{\beta}h_{\sigma\lambda})
+ 4 (\partial_{\rho}h^{\sigma\alpha}) (\partial_{\sigma}h^{\rho\beta})
\nonumber \\ 
+ 2 (\partial^{\rho}h^{\alpha\beta}) (\partial_{\rho}h) 
- 4 (\partial^{\rho}h^{\sigma\alpha}) (\partial_{\rho}h^{\beta}_{\cdot\sigma})
+ 2 (\partial^{\alpha}u_{\rho}) (\partial^{\beta}\tilde{u}^{\rho})
+ 2 (\partial^{\beta}u_{\rho}) (\partial^{\alpha}\tilde{u}^{\rho})
\nonumber \\
- 4 (\partial^{\alpha}v_{\rho}) (\partial^{\beta}v^{\rho})
+ 2 m^{2} \Bigl( 2 v^{\rho\alpha} h^{\beta}_{\cdot\rho} 
- v^{\alpha\beta} h \Bigl)
\eea
and
\bea
D_{2} = - {1\over 2} (\partial^{\rho}h) (\partial_{\rho}h)
+ (\partial^{\rho}h^{\alpha\beta}) (\partial_{\rho}h_{\alpha\beta})
- 2 (\partial^{\alpha}h^{\beta\lambda}) (\partial_{\beta}h_{\alpha\lambda})
- 2 (\partial^{\rho}u^{\sigma}) (\partial_{\rho}\tilde{u}_{\sigma})
 \Bigl]
\nonumber \\
+ 2 (\partial^{\rho}v^{\sigma}) (\partial_{\rho}v_{\sigma})
+ m^{2} \Bigl( - h^{\alpha\beta} h_{\alpha\beta}
+ {1\over 2} h^{2} \Bigl),
\eea
\be
D_{2}^{\alpha\beta;\nu} = {\cal D}_{2}^{\alpha\beta;\nu} 
+ \eta^{\alpha\beta} D^{\nu}_{2}
\ee
with
\bea
{\cal D}_{2}^{\alpha\beta;\nu} = 
- 4 (\partial_{\rho}h^{\alpha\beta}) h^{\nu\rho}
- 4 (\partial^{\alpha}h^{\nu\sigma}) h^{\beta}{\cdot\sigma}
- 4 (\partial^{\beta}h^{\nu\sigma}) h^{\alpha}{\cdot\sigma}
\nonumber \\
- 2 h^{\alpha\beta} (\partial^{\nu}h) 
+ 4 h^{\alpha\rho} (\partial^{\nu}h^{\beta}_{\cdot\rho})
+ 4 h^{\beta\rho} (\partial^{\nu}h^{\alpha}_{\cdot\rho})
+ 2 \eta^{\alpha\nu} u_{\rho} (\partial^{\rho}\tilde{u}^{\beta})
+ 2 
\eta^{\beta\nu} u_{\rho} (\partial^{\rho}\tilde{u}^{\alpha})
\eea
\bea
D_{2}^{\nu} = 4 (\partial^{\rho}h^{\nu\sigma}) h_{\rho\sigma}
+ h (\partial^{\nu}h)
- 2 h_{\rho\sigma} (\partial^{\nu}h^{\rho\sigma})
- 2 u_{\rho} (\partial^{\rho}\tilde{u}^{\nu}) \Bigl]
\eea

\be
D_{3}^{\rho} = - 4 \Bigl[ (\partial^{\sigma}h_{\lambda\sigma}) 
(\partial^{\rho}\tilde{u}^{\lambda})
- m (\partial^{\rho}v^{\sigma}) 
\tilde{u}^{\sigma} \Bigl]
\ee
\be
D_{3}^{\rho;\nu} = - 4 h^{\nu\sigma} (\partial_{\sigma}\tilde{u}^{\rho})
\ee

\be
D_{4}^{\rho} \equiv 4 m (\partial^{\sigma}v^{\rho}) u_{\sigma}
\ee
\be
D_{4}^{\rho;\nu} \equiv 4 \Bigl[ (\partial_{\sigma}h^{\rho\sigma}) u^{\nu}
- h^{\nu\sigma} (\partial_{\sigma}u^{\rho}) \Bigl]
\ee

\be
D_{5}^{\rho} \equiv 0
\ee
\be
D_{5}^{\rho;\nu} \equiv 2 \Bigl[ - 2 h^{\nu\sigma} (\partial_{\sigma}v^{\rho})
+ m u^{\nu} \tilde{u}^{\rho}
 \Bigl].
\ee

One also has:
\bea
[ h_{\mu\rho}(x), T(y) ] = i~\partial_{\mu}D_{m}(x-y) E_{\rho}(y)
 + \cdots
\nonumber \\
~[ \partial_{\nu}h_{\mu\rho}(x), T(y) ] 
= i~\partial_{\mu}\partial_{\nu}D_{m}(x-y) E_{\rho}(y)
 + \cdots
\eea
with
\be
E_{\rho} = 2 u^{\sigma} (\partial_{\sigma}\tilde{u}_{\rho}).
\ee

From these formul\ae~we obtain
\be
[ T^{\mu}(x), T(y) ] = i~\partial_{\mu}D_{m}(x-y) A(x,y)
+ i~\partial_{\mu}\partial_{\nu}D_{m}(x-y) A^{\nu}(x,y) 
+ \cdots
\label{ano1}
\ee
where
\be
A(x,y) \equiv T_{1}(x) {\cal D}_{1}(y) 
+  T_{2}^{\alpha\beta}(x) {\cal D}_{2;\alpha\beta}(y) 
+ \sum_{j=3}^{5}T_{j}^{\nu}(x) D_{j;\nu}(y)
 + S^{\rho}(x) E_{\rho}(y)
\ee
and
\be
A^{\nu}(x,y) \equiv T_{1}(x) {\cal D}_{1}^{\nu}(y) 
+ T_{2;\alpha\beta}(x) {\cal D}_{2}^{\alpha\beta;\nu}(y) 
\nonumber \\
+ \sum_{j=3}^{5} T_{j;\rho}(x) D_{j}^{\rho;\nu}(y) 
+ S^{\nu\rho}(x) E_{\rho}(y)
\ee
where
\bea
{\cal D}_{1} \equiv D_{1} - 2 D_{2} 
\nonumber \\
{\cal D}^{\nu}_{1} \equiv D^{\nu}_{1} - 2 D^{\nu}_{2}.
\eea

From formula (\ref{ano1}) we get the anomaly
\be
a(x,y) = \delta(x-y) A(x) 
+ [ \partial_{\nu}\delta(x-y) ] A^{\nu}(x,y)
\ee

By integration by parts we can rewrite this as follows:
\be
a(x,y) = \delta(x-y) {\cal A}(x) 
- \partial^{y}_{\nu} \Bigl[ \delta(x-y) {\cal A}^{\nu}(y) 
\Bigl]
\label{ano2}
\ee
with
\bea
{\cal A}(x) \equiv A(x) + \partial^{y}_{\nu}A^{\nu}(x,y)|_{y=x}
\nonumber \\
{\cal A}^{\nu}(x) \equiv 
A^{\nu}(x,x)
\label{ano3}
\eea

From the preceding formul\ae~ we get
\be
{\cal A} = T_{1} \tilde{\cal D}_{1} + 
T_{2;\alpha\beta} \tilde{\cal D}_{2}^{\alpha\beta} 
+ \sum_{j=3}^{5} T_{j;\rho} \tilde{D}_{j}^{\rho} 
+ S^{\rho} E_{\rho} + S^{\nu\rho} (\partial_{\nu}E_{\rho}) 
\label{total-ano}
\ee
where
\bea
\tilde{\cal D}_{j} \equiv {\cal D}_{j} + \partial_{\nu}{\cal D}_{j}^{\nu} 
\quad j = 1,2
\nonumber \\
\tilde{D}_{j} \equiv D_{j} + \partial_{\nu}D_{j}^{\nu} 
\quad j = 3,4,5,6
\eea

The total anomaly comes from the two commutators
$[ T^{\mu}(x), T(y) ] + [ T^{\mu}(y), T(x) ]$
and is
\be
{\cal A}(x,y) = a(x,y) + a(y,x).
\ee
Now, the second-order gauge invariance condition (\ref{gauge-n}) for
$n = 2$
is fulfilled {\it iff} one can write this anomaly as follows:
\be
{\cal A}(x,y) = 2 i d_{Q} N(x,y) + \partial_{\mu}^{x}N^{\mu}(x,y) 
+ \partial_{\mu}^{y}N^{\mu}(y,x) 
\ee
where
$N(x,y)$
and
$N^{\mu}(x,y)$
are quasi-local Wick polynomials. One can show rather easily that this 
condition 
is equivalent to:
\be
{\cal A} = i d_{Q} N + \partial_{\mu}N^{\mu}
\label{gauge2}
\ee
for some Wick polynomials (in one variable) $N$ and 
$N^{\mu}$
of ghost number $0$ and resp. $1$; here
${\cal A}$
is given by the first formula (\ref{ano3}). From power counting 
considerations (\ref{power}) one also
 has the limitations
\be
deg(N),~deg(N^{\mu}) \leq 6.
\ee

The condition (\ref{gauge2}) is the basic condition which we will investigate
from now on; it splits in distinct conditions for different
sectors.

a) We consider first the part
${\cal A}_{uhhh}$
of the anomaly which is linear in 
$u_{\mu}$
and tri-linear in the field
$h_{\alpha\beta}$.
From the expression of the anomaly we have
\be
{\cal A}_{uhhh} = u^{\mu} {\bf A}_{\mu} + (\partial_{\mu}u^{\mu}) {\bf A}
+ (\partial^{\nu}u^{\mu}) {\bf A}_{\mu\nu}
\label{uHHH}
\ee
where
\bea
{\bf A}_{\mu} \equiv - (\partial_{\mu}h)~ \tilde{\cal D}_{1} 
+ 2 (\partial_{\mu}h^{\alpha\beta})~ \tilde{\cal D}_{2;\alpha\beta}^{hh}
\nonumber \\
{\bf A} \equiv - h \tilde{\cal D}_{1} 
+ 2 h^{\alpha\beta} \tilde{\cal D}_{2;\alpha\beta}^{hh}
\nonumber \\
{\bf A}_{\mu\nu} \equiv 2 h_{\mu\nu} \tilde{\cal D}_{1} 
- 4 h_{\mu}^{\cdot\rho} \tilde{\cal D}_{2;\nu\rho}^{hh}.
\label{As}
\eea
By the symbol $hh$ as an index means that we consider only the bi-linear
part in 
$h_{\alpha\beta}$
of the corresponding expression. We want to find out if it is possible to write
\be
{\cal A}_{uhhh} = i d_{Q} N_{1} + \partial_{\mu}N^{\mu}_{1}
\label{gauge-u}
\ee
for some Wick polynomials
$N_{1}$
and
$N_{1}^{\mu}$
of ghost numbers $0$ and resp. $1$. We will not consider the generic form of
$N_{1}$
and
$\tilde{N}^{\mu}_{1}$
;
instead, we make the following ansatz
\be
\tilde{N}^{\mu}_{1} = u_{\nu}~t^{\mu\nu} 
+ (\partial_{\rho}u_{\nu})~t^{\mu\nu\rho} 
\ee
where the expressions
$t^{\cdots}$
are tri-linear in the fields
$h_{\alpha\beta}$,
have null ghost number and are limited by the power counting conditions
\be
deg(t^{\mu\nu}
) \leq  5, \quad deg(t^{\mu\nu\rho}
) \leq 4;
\ee
we will also suppose that the expression
$t^{\mu\nu\rho}$
does not contain terms with the factor
$\eta^{\mu\rho}$.

If we consider in
$N_{1}$
only terms of the type
$hh (\partial h) (\partial h)$
and
$hhhh$,
we have the following generic expressions of
$d_{Q} N_{1}$:
\be
i~d_{Q} N_{1} = (\partial^{\nu}u^{\mu})~{\cal B}_{\mu\nu} 
+ (\partial^{\nu}\partial^{\rho}u^{\mu})~{\cal B}_{\mu\nu\rho}
\ee
where the expressions
${\cal B}_{\cdots}$
are tri-linear in the fields
$h_{\alpha\beta}$,
have null ghost number and are limited by the power counting conditions
\be
deg({\cal B}_{\mu\nu}
) \leq  5, \quad deg({\cal B}_{\mu\nu\rho}
) \leq 4;
\ee
we can suppose that the expression
${\cal B}^{\mu\nu\rho}$
is symmetric in 
$\nu, \rho$.

The explicit expressions for
${\cal B}_{\dots}$
follow from the generic ansatz for
$N_{1}$;
they depend on some unknown coefficients which are to be determined.
We substitute everything in the equation (\ref{gauge-u}) and get
 the following
system:
\bea
{\bf A}^{\mu} = \partial_{\nu}t^{\nu\mu}
\nonumber \\
~{\bf A}^{\mu\nu} + \eta^{\mu\nu} {\bf A} = {\cal B}^{\mu\nu}
+ t^{\nu\mu} + \partial_{\rho}t^{\rho\mu\nu}
\nonumber \\
~{\cal B}^{\mu\nu\rho} + {\cal S}_{\nu\rho} t^{\nu\mu\rho} = 0 
\label{gauge-u-ab}
\eea

From the last equation we can determine
$t^{\mu\nu\rho}$
(if we suppose that it is symmetric in $\mu$ and $\rho$). We substitute in the
second equation of the system and determine
$t^{\mu\nu}$.
Finally we substitute this in the first equation of the system; it turns out 
that the system is consistent {\it iff} the following equation is true:
\be
{\bf A}^{\mu} - \partial_{\nu}{\bf A}^{\mu\nu} - \partial^{\mu} {\bf A}
= - \partial_{\nu}{\cal B}^{\mu\nu} 
+ \partial_{\nu}\partial_{\rho}{\cal B}^{\mu\nu\rho}.
\label{basic} 
\ee

This is the basic equation which we now analyze. The expressions from the 
left hand side can be computed explicitly from the formul\ae~(\ref{As}). 
The general strategy is to make an ansatz for
$N_{1}$,
compute the expressions
${\cal B}_{\dots}$
and compute the right hand side of the equation. Then we get a system for the 
unknown coefficients of the finite renormalization
$N_{1}$.

The ansatz for 
$N_{1}$
can be guessed analyzing different contributions appearing in
${\cal A}_{uhhh}$
and it is
\bea
N_{1} = f_{1} h^{\mu\nu} h^{\rho\sigma} 
(\partial_{\rho}h_{\mu\nu}) (\partial_{\sigma}h)
+ f_{2} h^{\mu\nu} h^{\rho\sigma} 
(\partial_{\alpha}h_{\mu\nu}) (\partial^{\alpha}h_{\rho\sigma})
\nonumber \\
+ f_{3} h^{\mu\nu} h_{\nu\rho} 
(\partial^{\alpha}h^{\rho\beta}) (\partial_{\beta}h_{\mu\alpha})
+ f_{4} h^{\mu\nu} h^{\rho\sigma} 
(\partial_{\mu}h^{\rho\alpha}) ({\partial_{\nu}h_{\sigma}}^{\cdot\alpha})
\nonumber \\
+ f_{5} h^{\mu\nu} h_{\nu\rho} 
(\partial^{\alpha}h_{\mu\beta}) (\partial_{\alpha}h^{\rho\beta})
+ f_{6} h^{\mu\nu} h^{\rho\sigma} 
(\partial_{\alpha}h_{\mu\rho}) (\partial^{\alpha}h_{\nu\sigma})
\nonumber \\
+ f_{7} h^{\mu\nu} h_{\nu\rho} 
({\partial^{\alpha}h_{\mu}}^{\cdot\rho}) (\partial_{\alpha}h)
+ f_{8} h^{4} + f_{9} h^{\mu\nu} h_{\mu\nu} h^{2}
\nonumber \\
+ f_{10} h^{\mu\nu} h_{\nu\rho} {h^{\rho}}_{\cdot\mu} H
+ f_{11} h^{\mu\nu} h_{\mu\nu} h^{\rho\sigma} h_{\rho\sigma}
+ f_{12} h^{\mu\nu} h_{\nu\rho} h_{\mu\sigma} h^{\rho\sigma};
\label{n1}
\eea
after performing the computation of
$d_{Q}N_{1}$
we see that the expression
${\cal B}_{\mu\nu\rho}$
does not have terms with the factor
$\eta_{\nu\rho}$
so
$t_{\mu\nu\rho} = - {\cal B}_{\nu\mu\rho}$
does not have terms with the factor
$\eta_{\mu\rho}$;
this is consistent with the ansatz we have made for
$t_{\mu\nu\rho}$.
The consistency equation (\ref{basic}) gives after long but straightforward
computations:
\bea
f_{1} = 8 \quad f_{2} = - 4 \quad f_{3} = f_{4} = - 16 \quad
f_{5} = 16 \quad f_{6} = 8 \quad f_{7} = - 8 
\nonumber \\
f_{8} = {1\over 12} m^{2} \quad f_{9} = - m^{2} \quad
f_{10} = {8\over 3} m^{2} \quad f_{11} = m^{2} \quad
f_{12} = - 4 m^{2}.
\eea

b) We consider now the contribution
${\cal A}_{uu\tilde{u}h}$
of the anomaly which is tri-linear in the ghost fields and linear in the field
$h_{\alpha\beta}$. 
The explicit expression is:
\bea
{\cal A}_{uu\tilde{u}h} = 
8 \Bigl[
 u_{\lambda} (\partial^{\lambda}h^{\alpha\beta}) 
+ (\partial^{\lambda}u_{\lambda}) h^{\alpha\beta} 
- (\partial_{\lambda}u^{\alpha}) h^{\lambda\beta} 
- (\partial_{\lambda}u^{\beta}) h^{\lambda\alpha} \Bigl] 
\nonumber \\
\times \Bigl[
 (\partial_{\alpha}u_{\rho}) 
(\partial_{\beta}\tilde{u}^{\rho}) 
+ (\partial_{\alpha}u_{\rho}) (\partial^{\rho}\tilde{u}_{\beta})
+ u^{\rho} (\partial_{\alpha}\partial_{\rho}\tilde{u}_{\beta})\Bigl]
\nonumber \\
- 8 u_{\lambda} (\partial^{\lambda}u^{\nu})
\Bigl[ (\partial_{\sigma}h^{\rho\sigma}) 
(\partial_{\nu}\tilde{u}_{\rho}) 
+ (\partial_{\sigma}h^{\rho\sigma}) (\partial_{\rho}\tilde{u}_{\nu})
+ h^{\rho\sigma} (\partial_{\rho}\partial_{\sigma}\tilde{u}_{\nu})\Bigl]
\nonumber \\
- 8 u_{\lambda} (\partial^{\lambda}\tilde{u}^{\nu})
\Bigl[ (\partial^{\rho}\partial^{\sigma}h_{\nu\sigma}) u_{\rho} 
+ (\partial^{\sigma}h_{\nu\sigma}) (\partial_{\rho}u^{\rho})
- (\partial_{\rho}h^{\rho\sigma}) (\partial_{\sigma}u_{\nu})
- h^{\rho\sigma} (\partial_{\rho}\partial_{\sigma}u_{\nu}) \Bigl]
\nonumber \\
+ 8 (\partial_{\nu}u_{\lambda})
 (\partial^{\lambda}h^{\nu\rho}) u^{\sigma} 
(\partial_{\sigma}\tilde{u}_{\rho})
\nonumber \\
+ 8 \Bigl[
 - u^{\nu} (\partial_{\lambda}h^{\rho\lambda}) 
+ u_{\lambda} (\partial^{\lambda}h^{\nu\rho}) 
+ (\partial_{\lambda}u^{\rho}) h^{\nu\lambda} \Bigl] 
\times 
\Bigl[
 (\partial_{\nu}u^{\sigma}) 
(\partial_{\sigma}\tilde{u}_{\rho}) 
+ u^{\sigma} (\partial_{\nu}\partial_{\sigma}\tilde{u}_{\rho}) \Bigl].
\label{ghost-ano}
\eea

We want to write this expression in the form
\be
{\cal A}_{uu\tilde{u}h} = i (d_{Q} N_{2})_{uu\tilde{u}h} 
+ \partial_{\mu}N^{\mu}_{2}.
\label{gauge-uuu}
\ee

We make the ansatz
\be
N_{2} = f_{13} u^{\rho} (\partial_{\rho}\tilde{u}^{\nu}) 
u^{\sigma} (\partial_{\sigma}\tilde{u}_{\nu}).
\ee

It is convenient to re-write the preceding formula (using ``partial 
integration") as follows:
\be
{\cal A}_{uu\tilde{u}h} - i (d_{Q} N_{2})_{uu\tilde{u}h}  
= \partial_{\mu}N^{\mu}_{2} + h_{\alpha\beta} U^{\alpha\beta} 
\ee
where 
$U^{\alpha\beta}$ 
is tri-linear in the ghost fields. One can obtain after some computations
that one can fix the constant 
$f_{13}$ 
such that
$U^{\alpha\beta} = 0$;
namely we must have
\be
f_{13} = 16. 
\ee

c) The contribution linear in the field 
$v^{\mu}$
of the anomaly is
\be
{\cal A}_{v} = 8 m \Bigl[ u^{\sigma} (\partial_{\sigma}u_{\nu})
+ (\partial_{\sigma}u^{\sigma}) u_{\nu} \Bigl ]
(\partial^{\nu}v^{\rho}) \tilde{u}_{\rho}
.
\ee

We have to subtract from this expression the contribution linear in
$v_{\mu}$
from
$d_{Q}N_{2}$. 
The result is a divergence:
\be
{\cal A}_{v} = i (d_{Q} N_{2})_{v} + \partial_{\mu}N^{\mu}_{3} \qquad
N_{3}^{\mu} \equiv 
8 m u^{\mu} u^{\nu} (\partial_{\nu}v_{\rho}) \tilde{u}_{\rho}
.
\ee

d) Now we consider the contribution bi-linear in the field
$v_{\mu}$;
we have
\bea
{\cal A}_{vv} = - 8 \Bigl[
 u_{\rho} (\partial^{\rho}h^{\alpha\beta}) 
+ (\partial^{\rho}u_{\rho}) h^{\alpha\beta} 
- (\partial_{\rho}u^{\alpha}) h^{\rho\beta} 
- (\partial_{\rho}u^{\beta}) h^{\rho\alpha} \Bigl] 
(\partial_{\alpha}v_{\sigma}) (\partial_{\beta}v^{\sigma})
\nonumber \\
+ 16 u_{\rho} (\partial^{\rho}v^{\nu}) 
\Bigl[ (\partial_{\lambda}h^{\lambda\sigma}) (\partial_{\sigma}v_{\nu}) 
+ h^{\lambda\sigma} (\partial_{\lambda}\partial_{\sigma}v_{\nu}) \Bigl]. 
\eea

One can write this expression as a total divergence:
\bea
{\cal A}_{vv} = \partial_{\mu}N^{\mu}_{4}
\nonumber \\
N^{\mu}_{4}
 \equiv - 8 \Bigl[ u^{\mu} h^{\alpha\beta}
(\partial_{\alpha}v_{\rho}) (\partial_{\beta}v^{\rho}) 
- 2 u_{\rho} h^{\mu\nu} (\partial^{\rho}v^{\sigma}) (\partial_{\nu}v_{\sigma}) 
\Bigl].
\eea

This finished the proof of the theorem.
$\qed$

We now consider the general case (i.e. 
$b \not= 0$).
It is sufficient to consider the first two sectors. The sector
$uhhh$
modifies only the values of the coefficients 
$f_{j}$.
The anomaly
${\cal A}_{uhhh}$
acquires the following expression: in (\ref{uHHH}) + (\ref{As}) we have
\bea
\tilde{\cal D}_{1} = \cdots + {1\over 2} m^{2} b~
\Bigl( h_{\rho\sigma} h^{\rho\sigma} - {1\over 2} h^{2} \Bigl)
\nonumber \\
\tilde{\cal D}_{2}^{\alpha\beta;HH} = \cdots + m^{2} b~
\Bigl( h^{\alpha\rho} {h_{\rho}}^{\cdot\beta} - {1\over 2} h^{\alpha\beta} h 
\Bigl)
\eea
where by $\cdots$ we mean the corresponding expressions from the case
$a = 1, b = 0$
multiplied by
$a^{2}$.
If we perform the same computations as in the preceding theorem we obtain the
new values of the coefficients
$f_{j}$:
\bea
f_{1} = 8 a^{2} \quad f_{2} = - 4 a^{2} \quad f_{3} = f_{4} = - 16 a^{2} \quad
f_{5} = 16 a^{2} \quad f_{6} = 8 a^{2} \quad f_{7} = - 8 a^{2}
\nonumber \\
f_{8} = {1\over 12} \Bigl( a^{2} + {1\over 4} a b \Bigl) m^{2} \quad
f_{9} = - \Bigl( a^{2} + {1\over 4} a b \Bigl) m^{2} \quad
f_{10} = {8\over 3} \Bigl( a^{2} + {1\over 4} a b \Bigl) m^{2} \quad
\nonumber \\ 
f_{11} = \Bigl( a^{2} + {1\over 4} a b \Bigl) m^{2} \quad
f_{12} = - 4 \Bigl( a^{2} + {1\over 4} a b \Bigl) m^{2}.
\eea

The sector
$uu\tilde{u}h$
gives the following expression of the anomaly:
\be
{\cal A}_{uu\tilde{u}h} = \cdots +
m^{2} a b {\cal A}_{1} + {1\over 16} m^{2} b^{2} {\cal A}_{2}
\ee
where
\bea
{\cal A}_{1} \equiv
{1\over 2} \Bigl[ -
 u_{\lambda} (\partial^{\lambda}h) 
- (\partial^{\lambda}u_{\lambda}) h 
+ 2 (\partial_{\lambda}u_{\nu}) h^{\lambda\nu} \Bigl ] 
u^{\rho} \tilde{u}_{\rho} 
\nonumber \\
+ 2 \Bigl[ u_{\lambda} (\partial^{\lambda}h^{\alpha\beta}) 
+ (\partial^{\lambda}u_{\lambda}) h^{\alpha\beta} 
- (\partial_{\lambda}u^{\beta}) h^{\lambda\alpha}
- (\partial_{\lambda}u^{\alpha}) h^{\lambda\beta} \Bigl] 
u_{\alpha} \tilde{u}_{\beta}
\nonumber \\
+ 2 u_{\lambda} (\partial^{\lambda}u^{\rho}) 
\Bigl( h_{\rho\sigma} \tilde{u}^{\sigma} 
- {1\over 4} h \tilde{u}^{\rho} \Bigl) 
- 2 u_{\lambda} (\partial^{\lambda}\tilde{u}^{\rho}) 
\Bigl( h_{\rho\sigma} u^{\sigma} - {1\over 4} h u^{\rho} \Bigl), 
\eea
\bea
{\cal A}_{2} \equiv h \Bigl[ 
u^{\rho} (\partial^{\lambda}u^{\lambda})
 \tilde{u}_{\rho} 
- 2 u^{\rho} (\partial_{\lambda}u_{\rho}) \tilde{u}^{\lambda})
- 2 u^{\rho} (\partial_{\rho}u_{\lambda}) \tilde{u}^{\lambda}) \Bigl]
\nonumber \\
- 2 h^{\rho\sigma} \Bigl[ 
u_{\sigma} (\partial^{\lambda}u^{\lambda})
 \tilde{u}_{\rho}
- 2 u_{\sigma} (\partial_{\lambda}u_{\rho}) \tilde{u}^{\lambda})
- 2 u_{\sigma} (\partial_{\rho}u_{\lambda}) \tilde{u}^{\lambda})
\Bigl],
\eea
where by $\cdots$ we mean the expression (\ref{ghost-ano}) multiplied by
$a^{2}$.
If we try to write the total anomaly as a total divergence $+$ coboundary, we
obtain in the end:
$b = 0$. 
\newpage

\section{Conclusions}

The evidence for a dark energy in our Universe motivates the
introduction of the cosmological constant $\Lambda$ into Einstein's equations.
The corresponding Einstein - Hilbert Lagrangian is given by
\be
L_{E} = - {2\over\kappa^2} \sqrt{-g} (R-2\Lambda),\quad
\kappa^{2} = 32 \pi G
\ee
where $G$ is Newton's constant and 
$g=\det (g_{\mu\nu})$. 
We want to
 expand this in powers of $\kappa$. For this purpose it is
convenient to use the so-called Goldberg variables 
\be
\tilde{g}^{\mu\nu} = \sqrt{-g} g^{\mu\nu},\quad
\tilde{g}_{\mu\nu} = (-g)^{-1/2} g_{\mu\nu}.
\ee
Now we write this metric tensor as
\be
\tilde{g}^{\mu\nu} = \eta^{\mu\nu} + \kappa h^{\mu\nu}
\ee
\bea
\tilde{g}_{\mu\nu} = \eta_{\mu\nu} - \kappa h_{\mu\nu}
+ \kappa^{2} h_{\mu\alpha} 
h^{\alpha}_{\cdot\nu} 
- \kappa^{3} h_{\mu\alpha} h^{\alpha\beta} h_{\beta\nu} 
+ \dots
\nonumber 
 \\
= \eta_{\mu\lambda} ( \delta_{\nu}^{\lambda} - \kappa h^{\lambda}_{\cdot\nu} 
+ \kappa^{2} 
h^{\lambda}_{\cdot\alpha} h^{\alpha}_{\cdot\nu}
- \kappa^{3} h^{\lambda}_{\cdot\alpha} h^{\alpha\beta} 
h_{\beta\nu} + \ldots )
\eea
Then we can expand the determinant $g$ and the Ricci scalar $R$ in (1)
in powers of $\kappa$; details can be found in \cite{Sc2}, Sect.5.5.

The first four orders in $\kappa$, i.e. $O(\kappa^{-2},\ldots
O(\kappa^2)$ come out to be:
\bea
L_{E} = {4\Lambda\over\kappa^2} 
+ {2\over\kappa}(\partial^{2}h - 
\partial_{\alpha}\partial_{\beta}h^{\alpha\beta} + \Lambda h)
\nonumber \\
+ {1\over 2} (\partial_{\gamma}h_{\alpha\beta}) 
(\partial^{\gamma}h_{\alpha\beta}) 
- {1\over 2} \partial_{\gamma}h
) (\partial^{\gamma}h)
+ (\partial_{\alpha}h^{\alpha\beta}) (\partial_{\beta}h) 
- (\partial_{\gamma}h^{\alpha\beta}) (\partial_{\alpha}h_{\beta}^{\cdot\gamma})
\nonumber \\
+ \Lambda \Bigl( {1\over 2}  h^{2} - h^{\alpha\beta} h_{\alpha\beta} \Bigl)
\nonumber \\
+
 \kappa \Bigl[ L^{(1)} + 4 \Lambda \Bigl( 
{1\over 6} h^{\alpha\beta} h_{\beta
\gamma} h^{\gamma}_{\cdot\alpha}
- {1\over 8} h^{\alpha\beta} h_{\alpha\beta} h
+ {1\over 48} h^{3} \Bigl) \Bigl]
\nonumber \\ 
+ \kappa^{2} \Bigl\{ L^{(2)} + 4 \Lambda \Bigl[ 
{1\over 32} \Bigl( h^{\alpha\beta} 
h_{\alpha\beta} \Bigl)^{2} 
+ {1\over 12} h h^{\alpha\beta} h_{\beta\gamma}h^{\gamma}
_{\cdot\alpha}
- {1\over 32} h^{2} h^{\alpha\beta} h_{\alpha\beta}
+ {1\over 4!} 
{h^4\over 16}
- {1\over 8} h^{\alpha\beta} h_{\beta\gamma} h^{\gamma\nu} h_{\nu\alpha} 
\Bigl] \Bigl\}
\label{expansion}
\eea
Here the terms 
$L^{(1)}$ 
and 
$L^{(2)}$ 
without $\Lambda$ are the same as in the 
ordinary massless gravity:
\bea
L^{(1)} \equiv - {1\over 4} h^{\alpha\beta} 
(\partial_{\alpha}h) (\partial_{\beta}h)
+ {1\over 2} 
h^{\mu\nu} (\partial_{\mu}h^{\alpha\beta})
(\partial_{\nu}h^{\alpha\beta}) 
+ h^{\alpha\beta} 
(\partial_{\nu}h^{\alpha\mu}) (\partial_{\mu}h^{\beta\nu})
\nonumber \\ 
+ {1\over 2} h^{\mu\nu} (\partial_{\alpha}h^{\mu\nu}) (\partial_{\alpha}h)
- h^{\mu\nu} (\partial_{\rho}
h^{\alpha\mu}) (\partial^{\rho}h^{\alpha\nu})
\eea
\bea
L^{(2)} \equiv 
- h_{\alpha\rho} h^{\rho}_{\cdot\beta} (\partial_{\nu}h^{\alpha\mu}) 
(\partial_{\mu}h^{\beta\nu})
- {1\over 2} h_{\rho\beta} h^{\beta}_{\cdot\gamma} 
(\partial_{\alpha}h^{\rho\gamma}) (\partial^{\alpha}h)
\nonumber \\
- {1\over 4} h_{\mu\nu} (\partial_{\alpha}h^{\mu\nu}) h_{\rho\gamma}
(\partial^{\alpha}h^{\rho\gamma})
+
 {1\over 2} h_{\mu\nu} (\partial_{\alpha}h^{\mu\nu}) h^{\alpha\beta}
(\partial_{\beta}h)
-
 h_{\alpha\rho} (\partial_{\mu}h^{\rho}_{\cdot\gamma}) 
(\partial_{\nu}h^{\alpha\gamma}) h^{\mu\nu}
\nonumber \\
+ h_{\rho\beta} h^{\beta}_{\cdot\gamma} (\partial_{\mu}h^{\alpha\rho}) 
(\partial_{\alpha}h^{\gamma}_{\cdot\alpha})
+
 {1\over 2} h_{\alpha\rho} h_{\beta\gamma} (\partial_{\mu}h^{\alpha\gamma})
(\partial^{\mu}h^{\beta\rho}).
\eea
The terms 
$O(\kappa^{-2})$ 
and 
$O(\kappa^{-1})$ 
in the Lagrangian (\ref{expansion}) 
are irrelevant because they give no 
equation of motion. The term 
$O(\kappa^{0})$ 
and quadratic in $h$
 
\bea
L^{(0)} \equiv 
{1\over 2} (\partial^{\mu}h^{\alpha\beta}) (\partial_{\mu}h_{\alpha\beta}
)
- (\partial_{\mu}h^{\alpha\beta}) (\partial_{\beta}h^{\mu}_{\cdot\alpha})
- {1\over 4} (\partial_{\alpha}h) (\partial^{\alpha}h)
+ \Lambda \Bigl( {h^2\over 2} - h^{\alpha\beta} h_{\alpha\beta}
 \Bigl)
\eea
gives the following Euler-Lagrange equation  
\bea
\partial^{2}h^{\alpha\beta} - \partial_{\mu} 
(\partial^{\beta}h^{\alpha\mu} + \partial^{\alpha}h^{\beta\mu})
- {1\over 2} \eta^{\alpha\beta} \partial^{2}h
- \Lambda (\eta^{\alpha\beta} h- 2 h^{\alpha\beta}) = 0.
\label{motion}
\eea
Taking the trace we find
\be
\partial_{\mu}\partial_{\alpha}h^{\alpha\mu} 
= - {1\over 2} \partial^{2}h - \Lambda h.
\label{constraint}
\ee
Differentiating (\ref{motion}) by 
$\partial_{\alpha}$ 
and substituting (\ref{constraint}) we derive
 the Hilbert gauge condition
\be
\partial_{\alpha}h^{\alpha\beta} = 0.
\ee
Then (\ref{constraint}) reduces to the Klein-Gordon equation 
\be
\partial^{2}h + 2 \Lambda h = 0
\ee
and from (\ref{motion}) we obtain the Klein-Gordon equation for the tensor 
field
\be
\partial^{2}h^{\alpha\beta} + 2 \Lambda h^{\alpha\beta} = 0
\ee
This means the graviton becomes massive with mass
\be
m^{2} = 2 \Lambda.
\label{mass}
\ee
Taking the current value of $\Lambda$ one finds a tiny mass 
($\approx
 10^{-32}$ eV) 
for the graviton, so that our massive theory passes all
 direct tests of 
general relativity. However, gravitational radiation
 requires a thorough 
investigation.

The cubic part 
$O(\kappa^1)$ 
in (\ref{expansion}) gives the coupling. Using (\ref{mass}) this agrees 
exactly with the pure graviton coupling terms in (\ref{first-order}), 
if we multiply with an overall factor 
$- 4$. 
Similarly, the pure graviton
 couplings in the quartic part 
$O(\kappa^2)$ 
agree with (\ref{second-order}) without the ghost terms, if we
 multiply by 16. 
This shows that our {\it massive gravity is the quantum
gauge theory corresponding to classical gravity with a cosmological
term.} Of course, the ghost couplings cannot be derived from the simple
Lagrangian above. To guess the right Lagrangian with ghost fields
(fermionic and bosonic) is not quite straightforward. The best way to
deduce the full theory is quantum gauge invariance, instead of some
classical heuristics.

In spin 1 gauge theories with massive gauge fields a Higgs field is
necessary to "generate" the masses. In the framework of quantum gauge
theory it is simply needed to restore gauge invariance to second order
\cite{Sc2}; first order gauge invariance holds without the Higgs couplings. 
It was a surprise for us when we found that massive gravity is gauge
invariant to first {\it and} second order without a Higgs field. In a way
this is even disappointing because a gravitational Higgs field would be
a nice candidate for the non-barionic dark matter.

A last remark is concerned with the mass zero limit of our theory.
As noticed above, in the limit 
$m\to 0$ 
the bosonic ghost 
$v^{\mu}$ 
does
 not completely decouple from the graviton. In fact, the term
$
4 h^{\mu\nu} (\partial_{\mu}v_{\rho})(\partial_{\nu}v^{\rho})
$ 
survives in (\ref{first-order}). That
 means the resulting massless theory is 
not identical with usual quantum
 gravity as discussed in \cite{Sc2}, for 
example. This leads to the conclusion
 that there exists at least two 
different quantum gauge theories which
 correspond to classical (massless) 
general relativity, one with an 
additional Bose field 
$v^{\mu}$ 
and one without.

\end{document}